\documentclass[review,3p,authoryear]{elsarticle}

\usepackage{amssymb}                                            
\usepackage{amsmath}
\usepackage{graphicx}
\usepackage{url}	
\usepackage[hidelinks]{hyperref} 

\usepackage{lineno}

\newcommand{\p}{\partial}
\newcommand{\f}[2]{\frac{#1}{#2}}

\hypersetup{
 colorlinks=true,
 allcolors=black,
 citecolor=blue
 }

\journal{Elsevier}

\begin{document}
\begin{frontmatter}

\title{Non-hydrostatic modelling of the wave-induced response of moored floating structures}
\author[add1,add2]{Dirk P. Rijnsdorp\corref{cor1}}

\ead{d.p.rijnsdorp@tudelft.nl}
\cortext[cor1]{Corresponding author.}

\author[add2]{Hugh Wolgamot}
\author[add1]{Marcel Zijlema}

\address[add1]{Environmental Fluid Mechanics section, Faculty of Civil Engineering and Geosciences, Delft University of Technology, Netherlands}
\address[add2]{The Oceans Graduate School, University of Western Australia, Australia}

\begin{linenomath*}
\begin{abstract}
Predictions of the wave-induced response of floating structures that are moored in a harbour or coastal waters require an accurate description of the (nonlinear) evolution of waves over variable bottom topography, the interactions of the waves with the structure, and the dynamics of the mooring system. In this paper, we present a new advanced numerical model to simulate the wave-induced response of a floating structure that is moored in an arbitrary nearshore region. The model is based on the non-hydrostatic approach, and implemented in the open-source model SWASH, which provides an efficient numerical framework to simulate the nonlinear wave evolution over variable bottom topographies. The model is extended with a solution to the rigid body equations (governing the motions of the floating structure) that is tightly coupled to the hydrodynamic equations (governing the water motion). The model was validated for two test cases that consider different floating structures of increasing geometrical complexity: a cylindrical geometry that is representative of a wave-energy-converter, and a vessel with a more complex shaped hull. A range of wave conditions were considered, varying from monochromatic to short-crested sea states. Model predictions of the excitation forces, added mass, radiation damping, and the wave-induced response agreed well with benchmark solutions to the potential flow equations. Besides the response to the primary wave (sea-swell) components, the model was also able to capture the second-order difference-frequency forcing and response of the moored vessel.  Importantly, the model captured the wave-induced response with a relatively coarse vertical resolution, allowing for applications at the scale of a realistic harbour or coastal region. The proposed model thereby provides a new tool to seamlessly simulate the nonlinear evolution of waves over complex bottom topography and the wave-induced response of a floating structure that is moored in coastal waters.
\end{abstract}
\end{linenomath*}


\end{frontmatter}

\section{Introduction}\label{sec:Introduction}
Accurate predictions of the wave-induced response of floating structures moored in coastal regions or harbours are important to ensure safe operations (e.g., loading and offloading of moored ships). Such predictions pose a challenging problem to numerical models due to the range of scales and physical phenomena involved. At local scales that span a few wave lengths, an accurate description of interactions between the waves and the floating structure (e.g., the scattering and radiation of waves) is required to accurately determine the wave forces that act on the structure. At large scales of typically many wave lengths, predicting the wave field in the vicinity of the floating structure requires an accurate description of the wave evolution from deep to shallower water in typically complex nearshore regions with variable bottom topography (e.g., due to the presence of entrance channels, breakwaters and quay walls). This includes linear wave processes like the refraction, diffraction, and reflection of waves. Furthermore, nonlinear interactions among sea-swell waves can also be significant in intermediate to shallow water and may transfer energy to relatively low wave frequencies \citep[e.g.,][]{LHS1962,Hasselmann1962}. Although these so-called infragravity waves are generally an order of magnitude smaller than the sea-swell waves, they have periods that may match the eigenperiod of a harbour and/or mooring system \citep[e.g.,][]{Bowers1977,Okihiro1993,Thotagamuwage2014,Cuomo2017InfragravityHarbor} and as a result may disrupt safe operations \citep[e.g.,][]{vanderMolen2006a,vanderMolen2016ImprovementHarbour}.  As an example, industry guidance for long-term moored nearshore structures \citep[e.g.,][]{DNVGL2019GuidanceSystems} explicitly recommends that wind-waves, infragravity waves and seiches all be considered in design.  An accurate description of the nonlinear wave field is thus critical for accurate predictions of the wave-induced response of floating structures in waters of restricted depth.

Over the past decades, various numerical methodologies have been developed to predict the wave-induced response of floating structures moored in coastal waters. Due to the range of scales involved, most existing models separately solve for the evolution of waves in the coastal region and the wave-structure interactions. With this approach, a wave propagation model accounts for the evolution of the waves as they propagate from deeper towards shallower water where the floating structure is moored. Subsequently, a hydrodynamic model that solves for the wave-structure interactions provides the hydrodynamic coefficients (i.e., the excitation forces, added mass, and radiation damping) that are subsequently used to solve the equations of motion that govern the wave-induced response of the moored floating structure.

Initial efforts used linear models to predict the wave-induced response of moored floating structures \citep[e.g.,][]{Oortmerssen1976,Sawaragi1982}. The local sea-state at the floating structure can, for example, be obtained from a model based on the mild-slope equations \citep{Berkhoff1972} that can simulate the linear evolution of a wave field in a complex harbour geometry \citep[e.g.,][]{Ohyama1994}. Assuming small waves and motions, a linear hydrodynamic model (e.g., based on the Boundary Element Method, BEM) can be combined with a frequency domain solution to the rigid-body equations to predict the wave-induced response of the moored floating structure  \citep[e.g.,][]{Kumar2016ModelingConditions}. However, the assumption of small waves and motions that allow for the use of linear theory can be overly restrictive as various forms of non-linearity can be important in such simulations.

For example, realistic mooring systems may introduce non-linearity even when motions are small \citep[e.g.,][]{Bingham2000}. As a result, time-domain solutions to the equation of motion \citep{Cummins1962} that account for the full non-linearity of the mooring system are generally preferred when computing the response of a moored floating structure. The excitation of infragravity waves by non-linear interactions (i.e., second-order difference-frequency interactions) among sea-swell waves introduces an additional source of non-linearity. For floating structures moored in open water, higher-order hydrodynamic models can be used to intrinsically account for the impact of infragravity waves on the wave-structure interactions \citep[e.g.,][]{You2015}. For more complicated coastal geometries such as a harbour, appropriate wave propagation models can be used to account for their impact. For example, \cite{vanderMolen2006a} used a dedicated infragravity wave model to predict the response of a moored ship-shaped vessel that was dominated by these low-frequency waves. Alternatively, Boussinesq-type wave models can provide the excitation force from both sea-swell and infragravity waves \citep{Bingham2000,vanderMolen2008}.

With this work, we propose an alternative method to predict the response of moored floating structures in realistic coastal and harbour regions. Our aim is to develop a single model that can seamlessly simulate the nonlinear evolution of waves in a complex nearshore region and predict the wave-induced response of a moored floating structure. Our numerical methodology is based on the non-hydrostatic approach. Non-hydrostatic wave-flow models were originally designed to simulate the evolution of (nonlinear) waves in coastal and oceanic waters \citep[e.g.,][]{Yamazaki2009,Zijlema2011,Ma2012,Cui2012}. Numerous studies have shown that non-hydrostatic models can accurately describe various relevant nearshore wave dynamics, including the steepening and eventual breaking of waves in the surf-zone \citep[e.g.,][]{Smit2013,Smit2014,Bradford2011,Derakhti2016a} and the excitation and dynamics of infragravity waves \citep[e.g.,][]{Rijnsdorp2014,Rijnsdorp2015a,DeBakker2016}. Furthermore, several studies have also shown that the non-hydrostatic framework can be extended to account for the wave-structure interactions with fixed (non-moving) floating obstacles \citep[e.g.,][]{Rijnsdorp2016,Ma2016,Ma2019ComparisonInteractions,Ai2019AMethod}. Importantly, non-hydrostatic models can describe these processes accurately while retaining computational efficiency due to the use of efficient numerical schemes \citep[e.g.,][]{Stelling2003} combined with describing the free-surface as a single-valued function of the horizontal coordinates, allowing for applications of the model at the spatial scales of a realistic coastal region \citep[e.g.,][]{Gomes2016,Risandi2020HydrodynamicModel,Rijnsdorp2021ASystem}.

In this paper, we present an extension to the non-hydrostatic wave-flow model SWASH \citep{Zijlema2011} to capture the wave-structure interactions and the wave-induced response of a moored floating structure. Our approach is based on the initial effort by \cite{Rijnsdorp2016}, who extended SWASH to account for the interactions between waves and a non-moving floating structure (i.e., the  diffraction problem). Here, we build on this methodology and extend the model with a solution to the rigid body equations that is tightly coupled to the hydrodynamic equations in order to capture the wave-induced response of the moored structure and to account for the radiation of waves by a moving structure (Section \ref{sec:NumericalMethodology}). We validated the proposed model for the wave-induced response of floating structures that are moored in open water (Section \ref{sec:TestCases}). Model predictions are compared with a frequency domain solution to the rigid body equations with hydrodynamic coefficients from models that solve the potential flow equations. We validate our model for the response to the primary (sea-swell) wave components for various wave conditions, ranging from monochromatic waves to short-crested sea states. To validate the model for the non-linear response, we also consider the second-order difference frequency loads and motions of a ship-shaped vessel that is moored in open water. In Section \ref{sec:Conclusions}, we summarize and discuss the findings of this work and conclude that the proposed model is able to capture the wave-induced response with a resolution that allows for model applications at the scale of a realistic harbour or coastal region.

\section{Numerical Methodology}\label{sec:NumericalMethodology}

\subsection{Governing equations}
The governing equations of the model are the Euler equations for an incompressible fluid of constant density. The fluid is bounded between the seabed $z=-d(x,y)$ at the bottom interface and the free-surface $z=\zeta(x,y,t)$ or a floating structure $z=S(x,y,t)$ at the top interface, in which $t$ is time and $\left( x,y,z \right)$ are the Cartesian coordinates. The governing equations read,

\begin{align}
    \f{\p u}{\p x} + \f{\p v}{\p y} + \f{\p w}{\p z}=0,\label{eq:LC}\\
    \f{\p u}{\p t} + u\f{\p u}{\p x} + v\f{\p u}{\p y} + w\f{\p u}{\p z} + g \f{\p \zeta}{\p x} + \f{\p p}{\p x} = 0,\label{eq:umom}\\
    \f{\p v}{\p t} + u\f{\p v }{\p x} + v\f{\p v }{\p y} + w \f{\p v}{\p z} + g \f{\p \zeta}{\p y} + \f{\p p}{\p y} = 0,\\
    \f{\p w}{\p t} + u \f{\p w }{\p x} + v \f{\p w }{\p y} + w\f{\p w}{\p z} + \f{\p p}{\p z} = 0,\label{eq:wmom}
\end{align}
in which $\bold{u}=(u,v,w)$ represents the velocity components in the Cartesian directions, $\zeta$ is the free-surface or piezometric head, $p$ is the non-hydrostatic pressure (normalised by the water density), and $g$ is the gravitational acceleration.

The vertical interfaces of the fluid are assumed to be a single valued function of the horizontal coordinates (i.e., the seabed, the free-surface and the hull of the structure). Here, the kinematic boundary conditions prescribe that the fluid velocity is equal to the material derivative of the respective interface. This results in the following kinematic boundary condition at the three interfaces,

\begin{align}
    w_{-d} = -u_{-d}\f{\p d}{\p x} -v_{-d}\f{\p d}{\p y},\\
    w_\zeta = \f{\p \zeta}{\p t} +u_\zeta\f{\p \zeta}{\p x} +v_\zeta\f{\p \zeta}{\p y},\\
    w_S = \f{\p S}{\p t} +u_S\f{\p S}{\p x} + v_S\f{\p S}{\p y}\label{eq:kbcS},
\end{align}
where the subscript indicates the $z-$location of the velocity components.
The velocities at the surface of the structure follow from the motions of the rigid body,

\begin{equation}
    \bold{u}_S = \f{\text{d} \bold{X}}{\text{d} t} + \f{\text{d} \Theta}{\text{d} t}\times\left(\bold{r}-\bold{r}_c\right),
\end{equation}
in which $\bold{X}=\left(X,Y,Z\right)$ and $\bold{\Theta}=\left(\theta_x,\theta_y,\theta_z\right)$ represent the translative and rotational motions of the floating structure, respectively, and $\bold{r}$ and $\bold{r_c}$ are the position vector on the wetted surface of the structure and the centre of gravity, respectively.

To close the set of equations that describe the fluid motion, a global continuity equation is derived by vertically integrating the local continuity equation (Eq. \ref{eq:LC}) and applying the relevant kinematic boundary conditions at the vertical interfaces that bound the fluid. In the region where the fluid is bounded by the water surface this results in,

\begin{equation}
    \f{\p \zeta}{\p t} + \f{\p}{\p x} \int_{-d}^{\zeta} u \text{d}z + \f{\p}{\p y} \int_{-d}^{\zeta} v \text{d}z = 0,\label{eq:GCa}
\end{equation}
and in the case the fluid is bounded by the structure this equation reads,

\begin{equation}
    \f{\p S}{\p t} + \f{\p}{\p x} \int_{-d}^{\zeta} u \text{d}z + \f{\p}{\p y} \int_{-d}^{\zeta} v \text{d}z = 0,\label{eq:GCb}
\end{equation}
where $\f{\p S}{\p t}$ follows from the kinematic boundary condition at the hull (Eq. \ref{eq:kbcS}).

The motions of a floating structure are described using the rigid body equations that follow from Newton's second law,

\begin{align}
    m\f{\text{d}^2 \bold{X}}{\text{d} t^2}=\bold{F},\label{eq:RB1}\\
    \bold{I}\f{\text{d}^2 \Theta}{\text{d} t^2} = \bold{M},\label{eq:RB2}
\end{align}
in which $m$ is the mass and $\bold{I}=\left(I_x, I_y, I_z\right)$ are the moments of inertia of the structure. $\bold{F}=\left( F_x, F_y, F_z\right)$ and $\bold{M}=\left( M_x, M_y, M_z\right)$ are all external forces and moments, respectively, that act on the body. In this work, we include the hydrodynamic loads and the forces from the mooring configuration. The hydrodynamic loads are computed by integrating the pressure over the surface of the body. The moorings considered in this work are modelled as linear spring-dampers,

\begin{equation}
    F_{M} = C + K l + B \f{\p l}{\p t},\label{eq:ML}
\end{equation}
where $C$ is a (constant) pretension, $K$ and $B$ are the spring and damping coefficient, respectively, and $l$ is the extension of the mooring line.

\subsection{Numerical implementation}

\begin{figure}[h]
    \centering
    \includegraphics{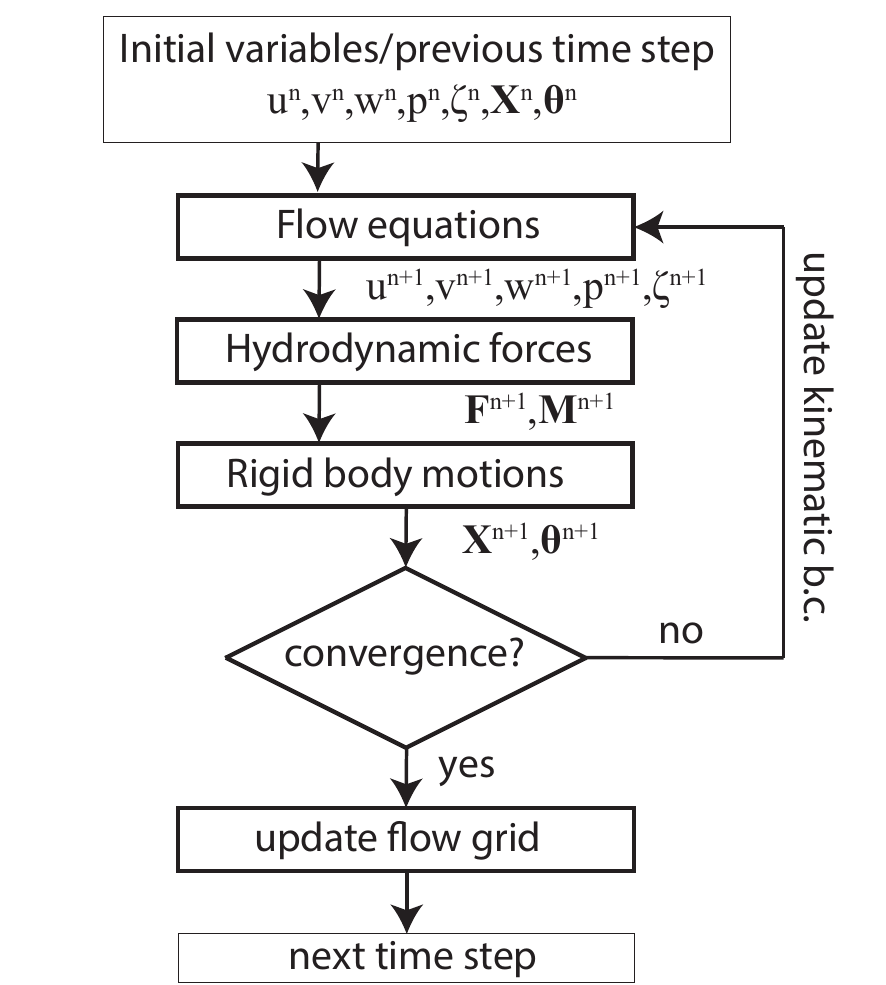}
    \caption{Flow diagram of the iterative procedure to solve the coupled fluid-structure interaction problem. The superscript of the variables indicates the time level, with $n$ indicating the current time step (or initial value) and $n+1$ at the next time step.}
    \label{fig:FlowDiagram}
\end{figure}

The resulting set of equations (Eqs. \ref{eq:LC}-\ref{eq:wmom} and Eqs. \ref{eq:RB1}-\ref{eq:RB2}) that describe the motion of the fluid and the floating structure represents a coupled problem. We adopt a partitioned approach to solve the governing equations, in which both sets of equations are treated independently. The coupling between the fluid and the structural equations is provided by the kinematic boundary conditions at the fluid-structure interface and the hydrodynamic forces that act on the structure. In this work, we use an implicit (or strongly coupled) approach to solve the fluid-structure interactions \citep[e.g.,][]{Matthies2003,Borazjani2008}. This implies that the kinematic boundary conditions and hydrodynamic forces depend on the implicit body motions and water pressure, respectively. To include these contributions a number of iterations are required during each time step until convergence is reached (see Fig. \ref{fig:FlowDiagram} for a flow diagram of the coupling algorithm). Convergence is reached when changes to the body motions and surface elevation within the iterative procedure are smaller than a user-defined tolerance $\epsilon$ (which was set to $\epsilon=1\times10^{-6}$ in all simulations of this work).

With the partitioned approach, both sets of equations are implemented in separate modules within the code. The flow equations are solved according to the general methods used in SWASH \citep[e.g.,][]{Zijlema2011}, and the rigid body equations are solved using an algorithm for simulating structural dynamics. In the following two sub sections, we present a brief description of the numerical methods that were used to solve both sets of equations.

\subsubsection{Hydrodynamic equations}
In SWASH, the equations governing the fluid motion are discretised on a curvilinear horizontal grid with a constant number of terrain following layers in the vertical. Flow variables are positioned on the grid based on a staggered arrangement \cite[See][for more details]{Zijlema2005,Zijlema2008,Zijlema2011}. To include the moving floating structure, we build on the implementation of \citet{Rijnsdorp2016} that accounted for the presence of a non-moving floating structure. In that work, the global continuity equations (Eqs. \ref{eq:GCa}-\ref{eq:GCb}) were recast into a single form following \citet{Casulli2013} to account for the pressurized flow underneath the structure and the wetting and drying of the hull. In combination with the semi-implicit $\theta$-method for the temporal discretisation, this allows us to capture the simultaneous occurrence of pressurized and free-surface flows. Further details regarding the temporal and spatial discretisation of the various terms can be found in \citet{Rijnsdorp2016}.

The pressure projection algorithm was used to account for the non-hydrostatic pressure \citep[e.g.,][]{Chorin1968,VanKan1986}. This method splits the time-integration into two parts: the hydrostatic and non-hydrostatic step. In the hydrostatic step, the global continuity equation (Eqs. \ref{eq:GCa}-\ref{eq:GCb}) is solved to compute the surface elevation and piezometric head, and an intermediate flow field. In the non-hydrostatic step the velocity field is corrected based on the non-hydrostatic pressure that is computed from the Poisson equation, which results from the substitution of the momentum equations (Eqs. \ref{eq:umom}-\ref{eq:wmom}) in the local continuity equation (Eq. \ref{eq:LC}). The global continuity equation is solved using a Newton-type iterative solver \citep{Brugnano2009}, and the Poisson equation is solved using a pre-conditioned BiCGSTAB solver.

With the terrain following grid system, the top layer follows the hull of the floating structure, and the kinematic boundary conditions at the hull (Eq. \ref{eq:kbcS}) are imposed directly at the fluid-structure interface. To keep the methodology simple and efficient, we do not account for the body motions in the flow grid. This implies that we linearize the kinematic boundary conditions at the hull of the floating body, analogous to, for example, linear potential flow theory and the BEM. With this approach however, the model does not intrinsically account for the hydrostatic restoring forces and moments. The hydrostatic restoring forces and moments were explicitly included as external forces based on analytical relations that account for linear changes in the buoyancy force due to the body motions \citep[e.g.,][]{Faltinsen1990}.

\subsubsection{Rigid body equations}
The equations describing the motions of the rigid body are solved using the generalized-$\alpha$ method \citep{Chung1993}, an implicit scheme that is based on the Newmark method \citep{Newmark1959}. For example, for an arbitrary displacement $X$ of a rigid floating body, the rigid body equation (Eq. \ref{eq:RB1}) is discretised as,

\begin{equation}
    m \left( \alpha_m A^n + (1-\alpha_m) A^{n+1} \right) = \alpha_f F^n + (1-\alpha_f) F^{n+1}.\label{eq:ga1}
\end{equation}
The displacement and the velocity of the floating body is subsequently computed as,
\begin{align}
    &X^{n+1} = X^n + \Delta t V^n + \Delta t^2\left(\left(1-\beta\right)A^n + \beta A^{n+1} \right),\\
    &V^{n+1} = V^n + \Delta t \left(\left(1-\gamma\right) A^{n} + \gamma A^{n+1} \right).
\end{align}

In these equations, $A$ and $V$ are the acceleration and velocity of the body; and $\alpha_f$, $\alpha_m$, $\beta$, and $\gamma$ are numerical parameters that control the accuracy and numerical dissipation of the method. The external force acting on the body $F$ includes the contributions from the hydrodynamics and mooring lines. The force in the mooring lines depend on the motions of the floating body through Eq. \ref{eq:ML}. To account for the implicit contributions in the mooring line force, we used an iterative approach to solve the generalized-$\alpha$ scheme. As mentioned before, the implicit contribution from the hydrodynamic forces and moments was included through several iterations over the flow and rigid body equations (as illustrated in Fig. \ref{fig:FlowDiagram}). In all simulations of this work that considered a moving floating structure, the numerical parameters were set at $\alpha_f=\alpha_m=\beta=0.5$, and $\gamma=0.25$.

\section{Test cases}\label{sec:TestCases}
The proposed model was validated for two floating structures of increasing complexity. In the first test case, we validated the model for the wave-induced response of a moored floating cylinder. This test case is representative of a point-absorber type wave-energy converter that is moored in coastal waters (with dimensions that are smaller than a typical wave length). In the second test case, we consider the wave-induced response of a ship-shaped vessel that is moored in open water of intermediate depth. The ship shape is a more complex hull form than the first test case, and has dimensions that are comparable to the typical wave length.

For both test cases, we validated the developed model for the diffraction problem, radiation problem, and the wave-induced response of the floating structure. The floating structures were subject to a range of sea states, varying from monochromatic waves to short-crested irregular waves, considering both the linear response to the primary wave field (Section \ref{sec:TC-FCOW}-\ref{sec:TC-MSOW}) and the nonlinear response associated with waves generated by second-order difference interactions (Section \ref{sec:TC-MSOW-O2}). Model predictions are compared to benchmark solutions to the potential flow problem; a semi-analytical solution for the moored floating cylinder and a BEM solution for the moored ship.





\subsection{Moored floating cylinder in open water}\label{sec:TC-FCOW}

This test case considers the interaction of waves with a floating cylinder, representative of a simplified wave-energy converter, that is located in water of 10 m depth. The cylinder had a radius of $10$ m, a height of $8$ m, a draft of $4$ m, and a density of $1/4$ of the water density ($\rho=1000$ kg/m$^3$). The cylinder was moored to the sea-bed with three tethers at a still-water depth of $10$ m. The tethers are equally distributed around the cylinder and point towards its centre of gravity with a vertical inclination of $45$ deg. The tethers included a constant pretension to counteract the positive buoyancy of the cylinder, with a spring coefficient equal to $5\times10^4$ N m$^{-1}$ and a damping coefficient of $5\times10^5$ Ns m$^{-1}$ (no attempt was made to optimise these coefficients to maximize power-take-off).

In the following, we compare model predictions with a benchmark solution based on the potential flow equations for the diffraction problem, the radiation problem, and the wave-induced response of the cylinder. The wave-induced response was computed based on a frequency domain solution to the linearized rigid body equations \citep[e.g.,][]{Sergiienko2018,Rijnsdorp2018}. The excitation forces and hydrodynamic coefficients (i.e., added mass and radiation damping) were obtained from an eigenfunction expansion solution to the linearized potential flow equations \citep[based on the work of][]{Jiang2014,Jiang2014a,Sergiienko2017,Sergiienko2018}.

\begin{figure}[h]
    \centering
    \includegraphics[scale=0.8]{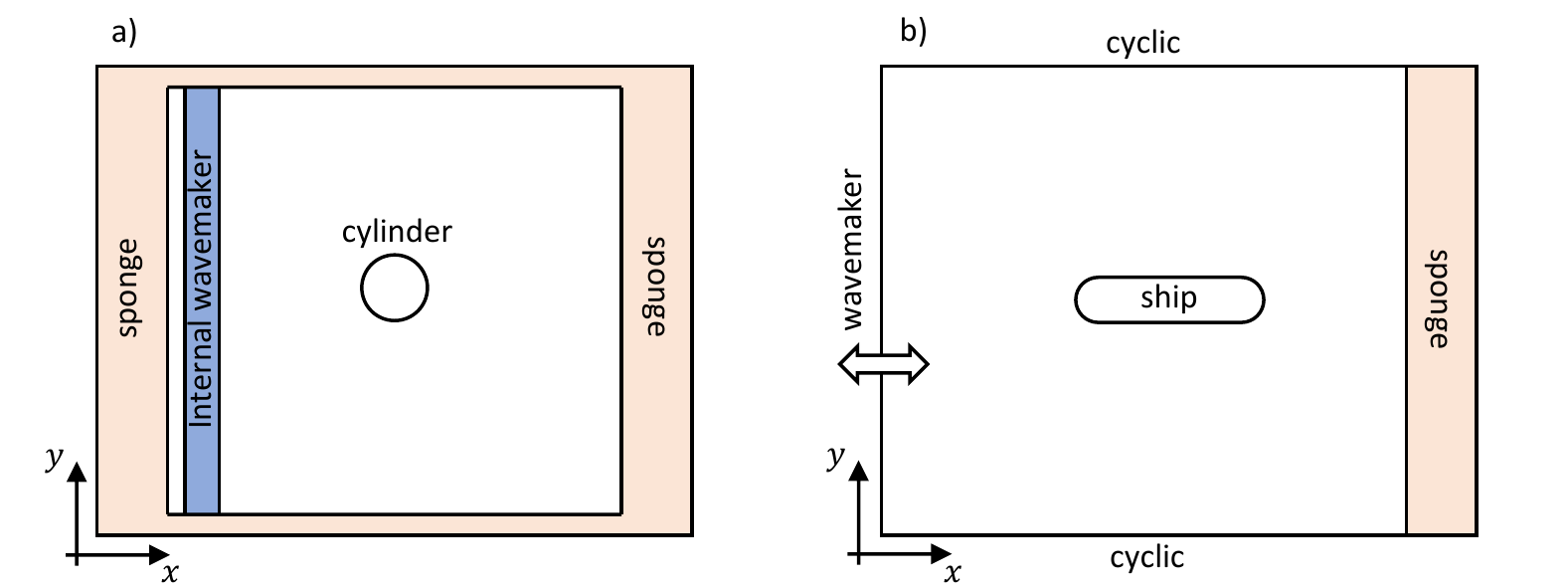}
    \caption{Sketch of the numerical domain used in the SWASH simulations for two example set-ups. Panel a) shows an example of the model-setup for a floating body (cylinder) that is subject to long-crested sea state, and panel b) shows the model-setup for a floating body (ship) that is subject to a short-crested sea-state.}
    \label{fig:NumDomain}
\end{figure}

In the SWASH simulations, the cylinder was located in a rectangular domain. The dimensions of the numerical domain were chosen to minimise the influence of side-wall reflections. For the long-crested diffraction and response simulations, the model domain spanned 700 m $\times$ 416 m. Waves were generated at the western boundary using a source function \citep{Vasarmidis2019InternalModel,Vasarmidis2020OnWaves}. To minimise the influence of re-reflections, sponge layers were positioned along all boundaries of the domain (see Fig. \ref{fig:NumDomain}a for an illustrative sketch). The sponge layers had a width of 150 m at the eastern and western end of the domain, and were 50 m wide along the northern and southern boundaries. For diffraction and response simulations with a short-crested sea state, we used a 700 $\times$ 2016 m domain with cyclic lateral boundaries and a weakly-reflective wavemaker at the western side of the domain (see Fig. \ref{fig:NumDomain}b for an illustrative sketch). For the radiation problem, the numerical domain spanned 350 $\times$ 350 m, with sponge layers of $150$ m at all lateral boundaries to absorb the waves radiated by the moving cylinder. With these model set-ups we found that the influence of side-wall reflections was minimal.

A rectilinear grid was used with a minimum horizontal grid resolution of $\Delta x=\Delta y = 2$ m that ensured at least 20 points per (peak) wave length and 10 points per cylinder diameter in the region of interest. Away from the region of interest, we used a coarser $\Delta y$ of 4 m to reduce the computational effort required. Two vertical layers were used in all cases, which was sufficient to capture the dispersion characteristics of the various sea-states. The time step was set at $1/200$ of the (peak) wave period.  For simulations with monochromatic waves, model results were analysed for 5 wave periods after spin-up. For the simulation with the irregular sea state, model results were analysed for a duration of 60 min after a spin-up time of 5 min.

\subsubsection{Monochromatic waves}\label{sec:TC-FCOW-MW}

We considered monochromatic waves with varying period ($T=4-14$ s with $1$ s increments) and a constant amplitude of $a=0.01$ m. In the following comparison, we only consider the loads and motions that are non-zero in the case of monochromatic incident waves (i.e., surge, heave and pitch). Model predictions were compared with the semi-analytical solutions for the exciting forces (the diffraction problem), the added mass and radiation damping (the radiation problem), and the wave-induced response of the floating body (the response amplitude operators (RAO) of the body).

\begin{figure}[b!]
    \noindent
    \makebox[\textwidth]{\includegraphics{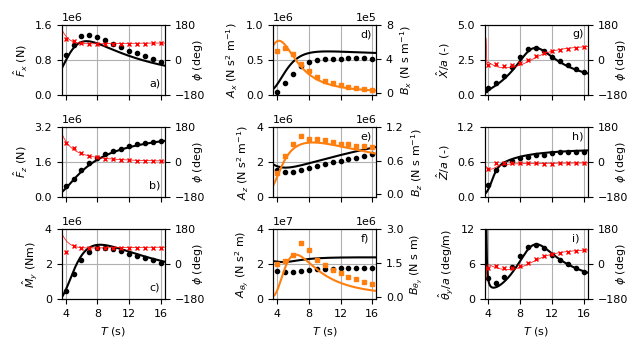}}
    \caption{Comparison between the SWASH predictions (markers) and semi-analytical potential flow solution (lines) for the diffraction problem (left panels), radiation problem (middle panels), and wave-induced response (right panels) of the floating cylinder. In panel a-c (diffraction problem) and panel g-i (wave-induced response), the black line and circles indicate the amplitude of the signal (left axis) and the red line and crosses indicate the phase difference with the incident wave signal (right axis). In panel d-f (radiation problem), the black line and circles indicate the added mass coefficient (left axis), and the orange line and squares indicate the radiation damping coefficient (right axis).}
    \label{fig:FC_DiffracResponse}
\end{figure}

First, we compare model predictions with the semi-analytical solution for the diffraction and radiation problem. Predictions of the amplitude of the excitation load agreed well for the range of wave periods considered (black line and markers in Fig. \ref{fig:FC_DiffracResponse}a-c), although the amplitude of the surge force $\hat{F}_x$ was slightly over predicted and the amplitude of the pitch moment $\hat{M}_y$ was slightly under predicted. The model also captured the phasing of the excitation load, except at the lowest wave period for $\hat{M}_y$ (red line and markers in Fig. \ref{fig:FC_DiffracResponse}a-c).

In Fig. \ref{fig:FC_DiffracResponse}d-f, we compare the hydrodynamic coefficients for the radiation problem (i.e., added mass and radiation damping). The hydrodynamic coefficients in SWASH were calculated based on the hydrodynamic loads from the radiation simulations (which did not include the hydrostatic restoring force as the flow grid was not updated based on the body motions). The hydrodynamic coefficients were computed from the load signals by taking the part of the hydrodynamic force that is in phase with the body velocity (radiation damping coefficient) and the part that is in phase with the body acceleration (added mass coefficient). Comparing the resulting coefficients with those from the semi-analytical solution to linear potential theory shows that SWASH captured their typical magnitude for all three degrees of freedom (Fig. \ref{fig:FC_DiffracResponse}d-f). Agreement was typically best for heave, and discrepancies were typically larger for surge and pitch.

To validate the model for the wave-induced response, we compared model predictions with a linear frequency domain solution based on the excitation forces (Fig. \ref{fig:FC_DiffracResponse}a-c) and hydrodynamic coefficients (Fig. \ref{fig:FC_DiffracResponse}d-f) from the semi-analytical LPF solution. In the SWASH simulations, the flow grid was not updated following the body motions (i.e., linearizing the kinematic boundary conditions on the hull), and the hydrostatic restoring forces and moments were computed based on the formulations of \cite{Faltinsen1990}.
The RAO and the phasing of all three degrees of freedom were found to be in excellent agreement with the reference solution (Fig. \ref{fig:FC_DiffracResponse}g-i). The model recovered the correct frequency dependent response, including an increased response in surge and pitch around $T\approx9$ s, and an increased response in pitch at shorter wave periods ($T<5$ s).

\subsubsection{Long-crested irregular waves}\label{sec:TC-FCOW-LCIW}
\begin{figure}[h]
    \noindent
    \makebox[\textwidth]{\includegraphics{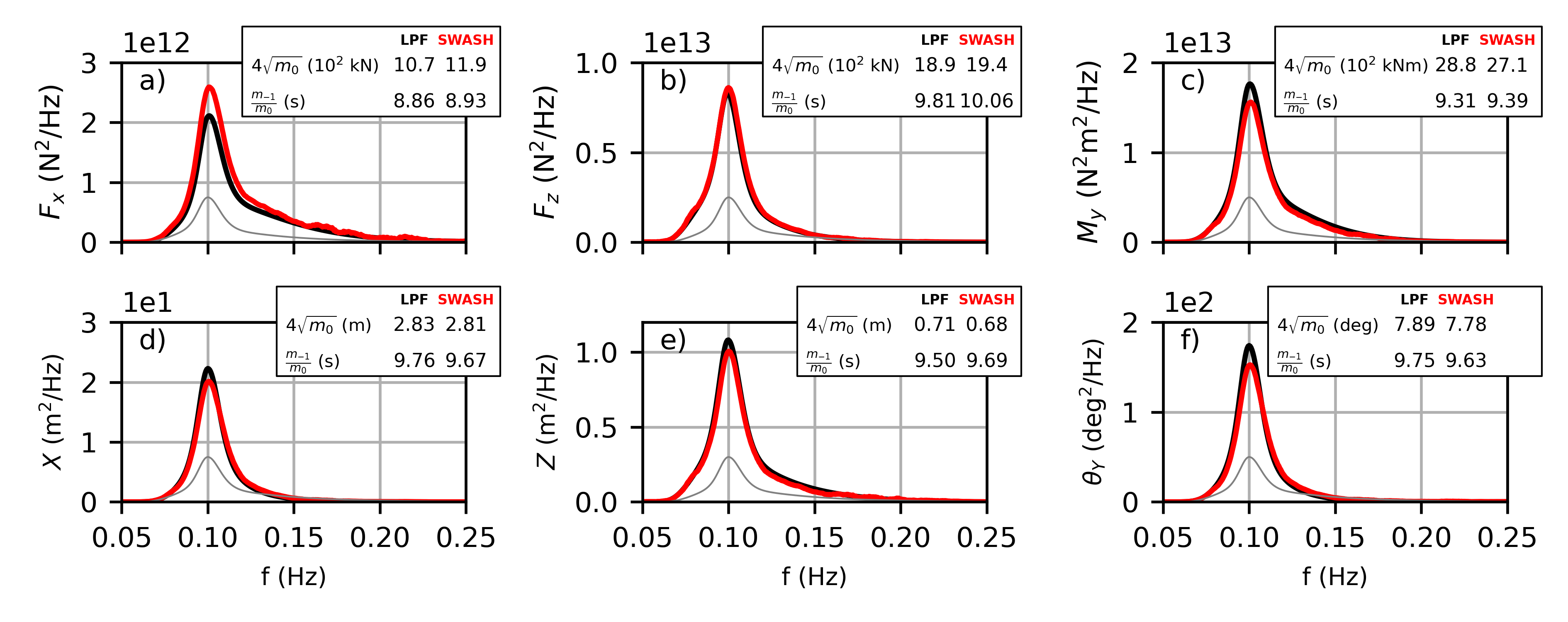}}
    \caption{Comparison between the SWASH predictions (red lines) and semi-analytical potential flow solution (black lines) for the excitation force (top panels), and wave-induced response (bottom panels) of the floating cylinder that is subject to a long-crested sea state. Bulk parameters based on spectral moments are included for each component in their respective panel. The thin gray line represents the incident wave spectrum.}
    \label{fig:FC_DiffRespJSWP}
\end{figure}

Next, we consider the excitation forces and wave-induced response for a long-crested irregular wave field. We considered irregular waves with a JONSWAP frequency distribution with a significant wave height of $H_s= 1$ m and a peak period of $T_p=10$ s. Fig. \ref{fig:FC_DiffRespJSWP} shows the power spectra components of the excitation forces and the (non-zero) components of the body motion. Power spectra were computed without smoothing, thereby providing a comparison for each wave frequency that was considered in the model. To allow for a quantitative comparison, representative bulk load or motions ($\sqrt{m_0}$) and mean period ($\sqrt{\frac{\textrm{m}_{-1}}{\textrm{m}_0}}$) were computed from the spectral moments $m_n$ (as indicated in each panel), with $m_n=\int f^n \text{PSD} \text{ d}f$ (in which $\text{PSD}$ is the power spectral density of the respective parameter).

The power spectra of the excitation forces were in good agreement with the LPF solution (Fig. \ref{fig:FC_DiffRespJSWP}a-c). Consistent with the results for monochromatic waves, $\hat{F}_x$ was slightly over estimated whereas $\hat{M}_y$ showed a slight underestimation. Differences between the bulk load ($4\sqrt{\textrm{m}_0}$) and mean load period ($\frac{m_{-1}}{m_0}$) were $<15\%$, confirming the general agreement between SWASH and LPF. The predicted response showed a comparable agreement for all three degrees of freedom, with differences between the bulk parameters $<5\%$ (Fig. \ref{fig:FC_DiffRespJSWP}d-f).

\subsubsection{Short-crested irregular waves}\label{sec:TC-FCOW-SCIW}
As a final scenario as part of this test case, we validated SWASH for a short-crested sea state. Short-crested irregular waves were generated for a JONSWAP spectrum (with $H_s = 1.0$ m and $T_p=10$ s) with a cosine directional distribution $\cos^m{\left(\theta\right)}$ with power $m=4$ (corresponding to a (half-width) directional spreading of $\approx25^\circ$). In SWASH, short-crested sea states are generated by means of a superposition of a large number of long-crested harmonic waves with a certain amplitude, frequency, phase, and direction \citep[See][for more details]{Rijnsdorp2015a}. To allow for a direct comparison with SWASH, the LPF based solutions were computed based on the target wave component with which the numerical wavemaker in SWASH was forced. Power spectra for all load and motion components were computed using Welch's method with fifteen $50\%$ overlapping windows.

\begin{figure}[h]
    \noindent
    \makebox[\textwidth]{\includegraphics{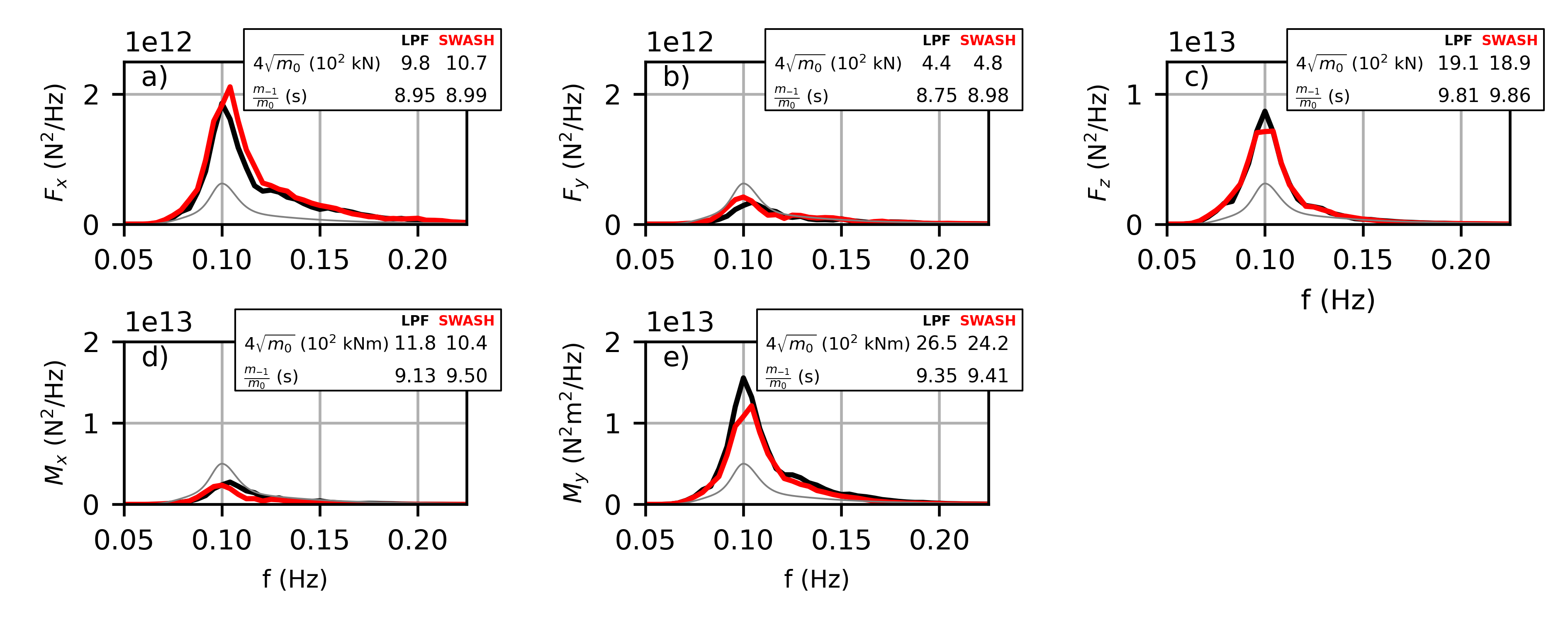}}
    \caption{Comparison between the SWASH predictions (red lines) and semi-analytical potential flow solution (black lines) for the excitation forces (panel a-c) and moments (panel d-e) of the floating cylinder that is subject to a short-crested sea state. Bulk parameters based on spectral moments are included for each force and moment component in their respective panel. The thin gray line represents the incident wave spectrum.}
    \label{fig:FC_DiffracJSWPsc}
\end{figure}

\begin{figure}[h!]
    \noindent
    \makebox[\textwidth]{\includegraphics{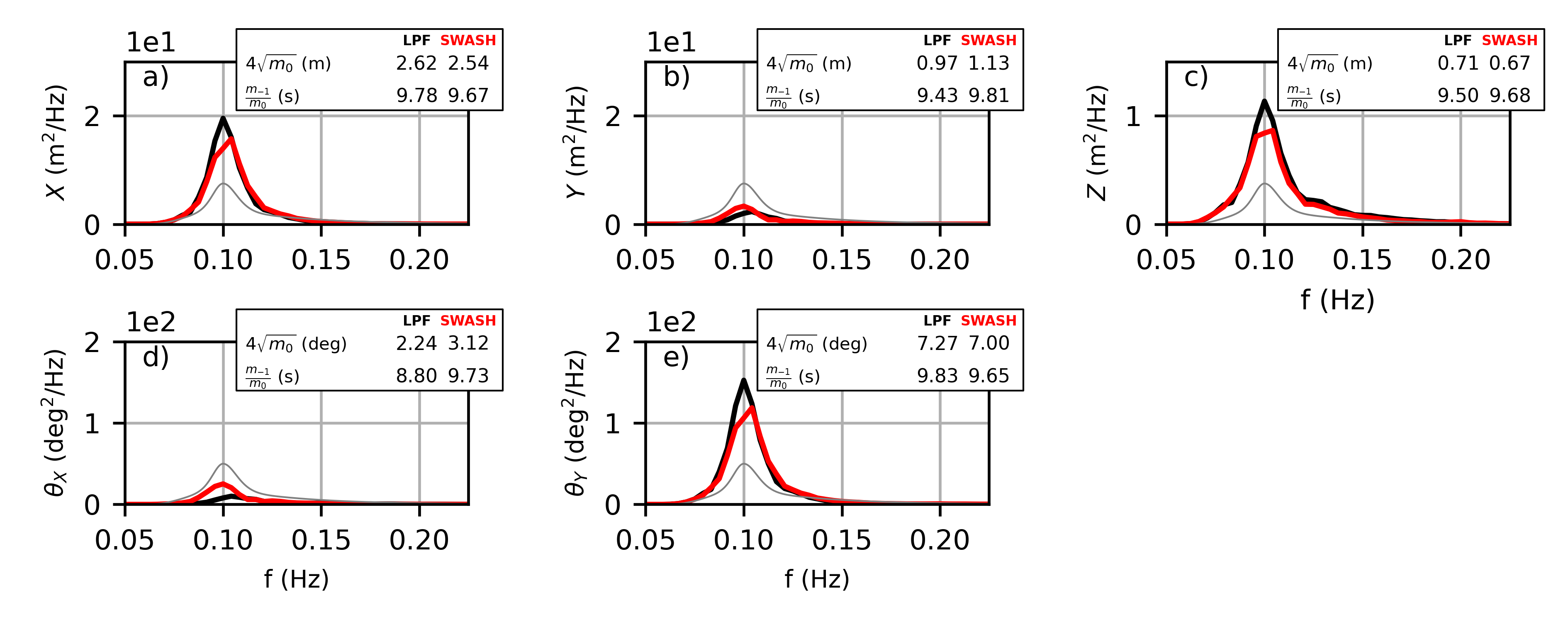}}
    \caption{Comparison between the SWASH predictions (red lines) and semi-analytical potential flow solution (black lines) for the wave-induced response of the floating cylinder that is subject to a short-crested sea state. Bulk parameters based on spectral moments are included for each body motion in their respective panel. The thin gray line represents the incident wave spectrum.}
    \label{fig:FC_RespJSWPsc}
\end{figure}

Due to the directional spreading of the wave field, the sway force and roll moment were non-zero (Fig. \ref{fig:FC_DiffracJSWPsc}b and \ref{fig:FC_DiffracJSWPsc}d), with a bulk magnitude that was approximately half of the surge force and pitch moment, respectively (Fig. \ref{fig:FC_DiffracJSWPsc}a and \ref{fig:FC_DiffracJSWPsc}e). The excitation forces and loads from SWASH and the linear potential flow solution were generally in good agreement for all five non-zero components, and the bulk load and mean load period parameters differed by $<15\%$ (Fig. \ref{fig:FC_DiffracJSWPsc}). A comparable agreement between models was found for the wave-induced response (Fig. \ref{fig:FC_RespJSWPsc}). Discrepancies between the two models were typically larger for sway and especially roll, for which differences between the bulk motions were $16\%$ and $38\%$, respectively.

\subsection{Moored ship in open water -- first order response}\label{sec:TC-MSOW}

\begin{figure}[!b]
    \centering
    \includegraphics{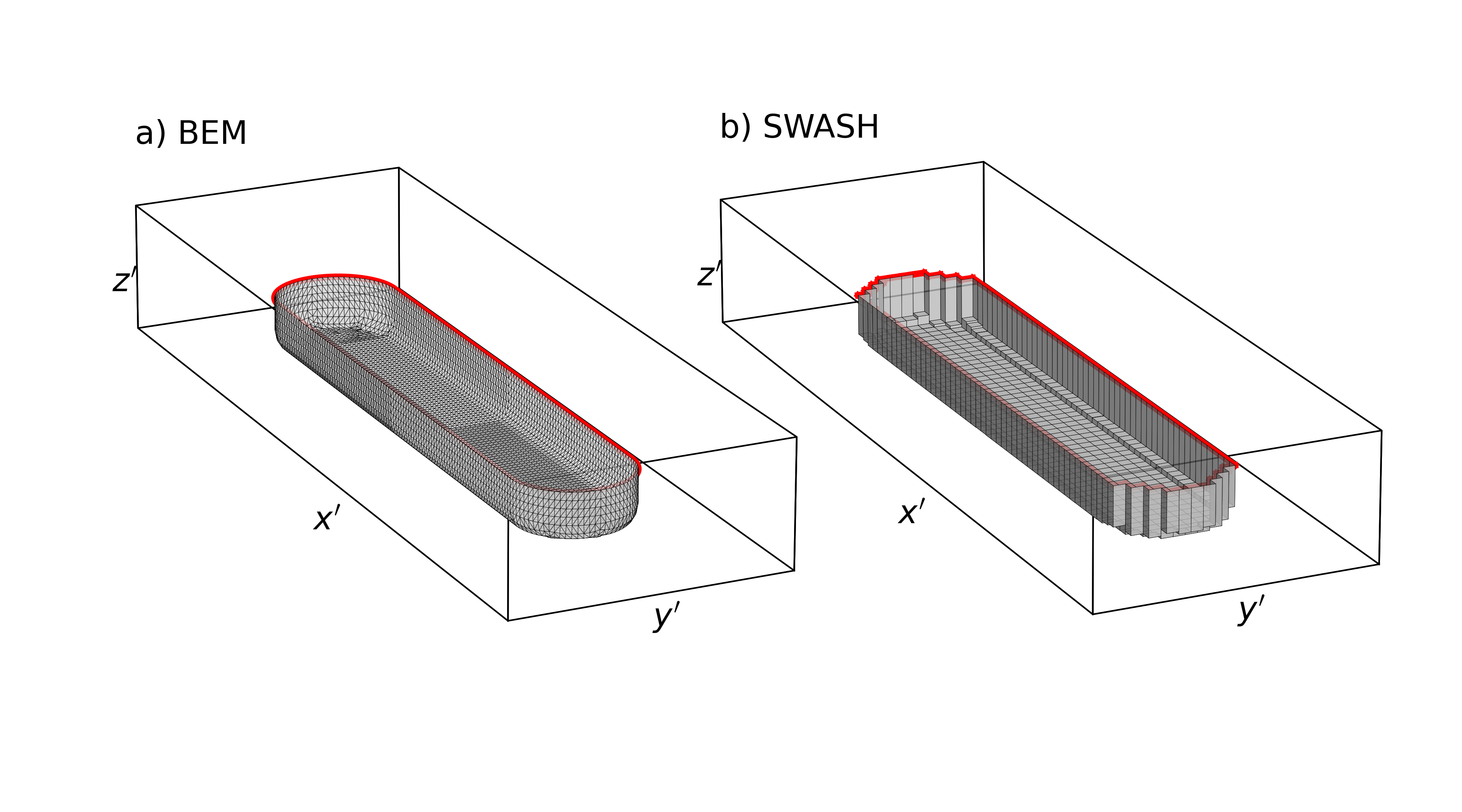}
    \caption{Sketch of the hull of the moored ship-shaped vessel used in the BEM (panel a) and SWASH (panel b) simulations. The thick red line in both panels indicate the waterline contour.}
    \label{fig:BEM_ShipHull}
\end{figure}

In the second test case, we consider the interactions of long-crested and short-crested head seas with a ship-shaped body that is moored in open water 28.6 m deep. SWASH predictions were compared with results from the BEM code DIFFRACT \citep[e.g.,][]{Sun2015WaveBarge,EatockTaylor1992WaveTheory}, (with a frequency-domain solution to the rigid body motions where necessary). For a linear calculation, the mean submerged hull of the vessel and the interior water-plane must be meshed. DIFFRACT employs isoparametric quadratic elements and the formulation employed avoids irregular frequencies. The main particulars of the hull are given in Table \ref{tab:Ship}, and Fig. \ref{fig:BEM_ShipHull} provides a sketch of the hull that was used in the BEM code (Fig. \ref{fig:BEM_ShipHull}a) and in SWASH (Fig. \ref{fig:BEM_ShipHull}b). The hull possessed two planes of symmetry and the vertical position of the centre of gravity of the vessel was located at $z=3.5$ m (i.e., above the mean free surface).

SWASH simulations for the diffraction problem and for the response to long-crested sea states were conducted in a domain that spanned 1850 $\times$ 1016 m in the $x$ and $y-$directions, respectively. Similar to the previous test case, waves were generated using a source function in combination with sponge layers to minimise the influence of re-reflections at the domain boundaries (see Fig. \ref{fig:NumDomain}a for an illustrative sketch). Sponge layers were positioned at the western and eastern boundaries (500 m wide), and the lateral (northern and southern) boundaries (100 m wide). For the radiation simulation, the domain spanned 1500 $\times$ 1100 m, with 500 m wide sponge layers along all lateral boundaries to absorb radiated waves. With this model set-up we found negligible influence from side-wall reflections. To accurately simulate the short-crested sea state, we used a larger basin (spanning 1850 $\times$ 5016 m in the $x$ and $y-$directions, respectively) with cyclic lateral boundaries and a weakly-reflective wavemaker at the western side of the domain (see Fig. \ref{fig:NumDomain}b for an illustrative sketch).

For all simulations, a rectilinear grid was used with a minimum horizontal grid resolution of $\Delta x=\Delta y = 4$ m that ensured at least 20 points per wave length in the region of interest and 10 cells along the vessels breadth. Two vertical layers were used, and the time step was set at $1/200$ of the (peak) wave period. For the regular wave simulations, model results were analysed for 10 wave periods after model spin-up. For the long-crested and short-crested wave simulations, model results were analysed for a duration of 60 min (240 peak wave periods) after model spin-up.

In all response simulations (of SWASH and the frequency-domain solution), a soft spring mooring ($K=609$ kN/m) was included in surge and sway to counteract the mean drift force. Due to the low stiffness, the motions of the moored body at the primary wave frequencies were not affected by the inclusion of this mooring (not shown).

\begin{table}
\begin{center}
\caption{Main particulars of the ship-shaped vessel}
\label{tab:Ship}
\begin{tabular}{c|c|c}
        & BEM & SWASH \\
     Length between perpendiculars $L_{pp}$ & 226 & 226 \\
     Breadth (m) & 43 & 43 \\
     Draft (m) & 11.5 & 11.5 \\
     Waterplane Area (m$^2$) & 9320.6 & 9440.3 \\
     Displacement (m$^3$) & 108100 & 104532
\end{tabular}
\end{center}
\end{table}

\subsubsection{Monochromatic waves}\label{sec:TC-MSOW-MW}

\begin{figure}[t]
    \noindent
    \makebox[\textwidth]{\includegraphics{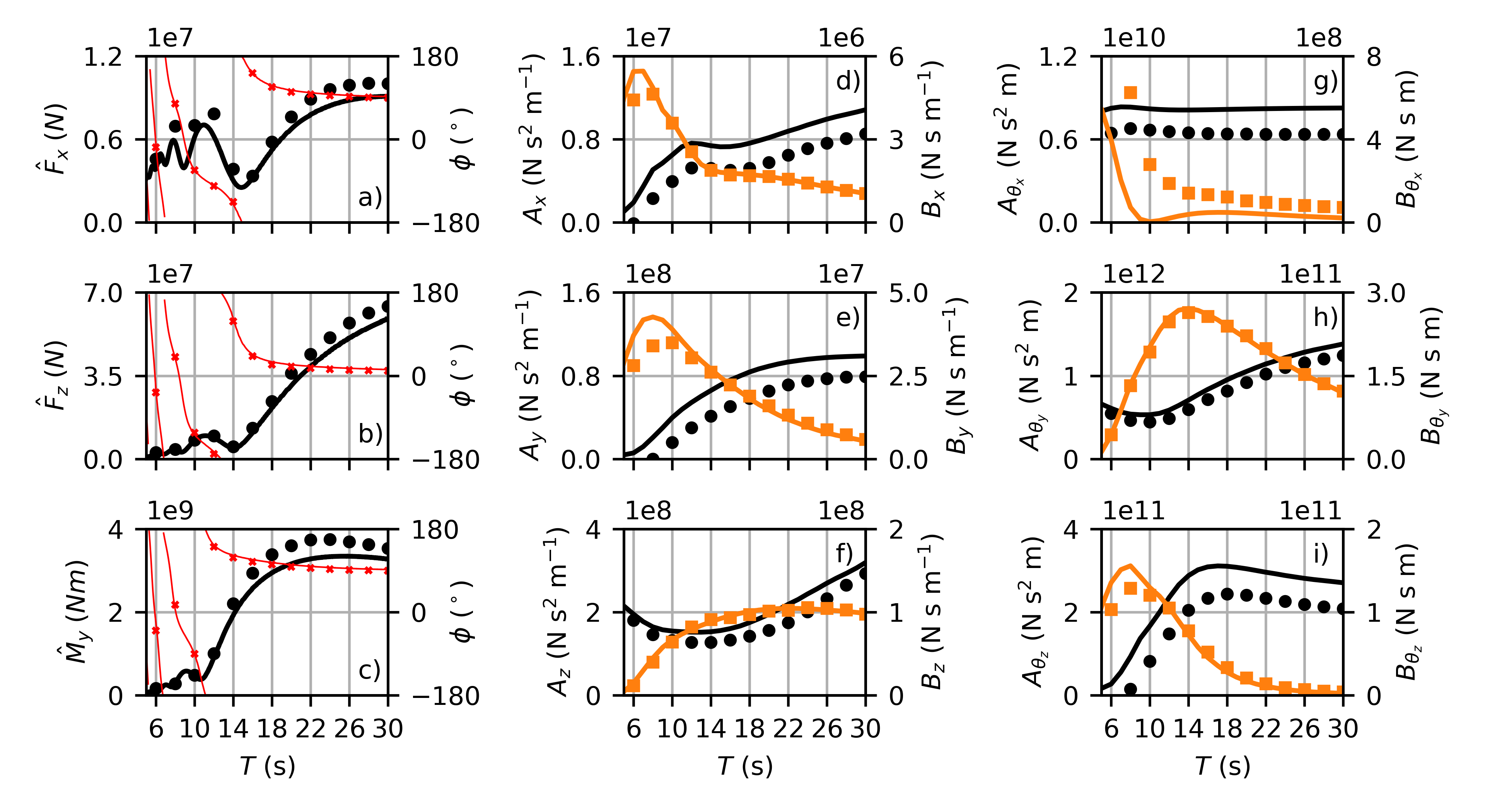}}
    \caption{Comparison between SWASH (markers) and BEM (lines) for the diffraction problem (panel a-c), and the radiation problem (panel d-h) of the moored ship-shaped vessel in open water. In panel a-c (diffraction problem), the black line and circles indicate the amplitude of the signal (left axis) and the red line and crosses indicate the phase difference with the incident wave signal (right axis). In panel d-i (radiation problem), the black line and circles indicate the added mass coefficient (left axis), and the orange line and squares indicate the radiation damping coefficient (right axis).}
    \label{fig:BEM_DifRad}
\end{figure}

The ship-shaped vessel was subject to a range of small amplitude ($a=0.01$ m) monochromatic waves with varying period ($T=6-30$ s with $2$ s increments). Fig. \ref{fig:BEM_DifRad} shows a comparison between the SWASH and BEM results for the diffraction and radiation problems. SWASH captured the typical magnitude and phase of the excitation forces ($\hat{F}_x$ and $\hat{F}_z$) and moment ($\hat{M}_y$) for the considered range of wave periods (Fig. \ref{fig:BEM_DifRad}a-c). Discrepancies between the two models were typically larger for the radiation problem (Fig. \ref{fig:BEM_DifRad}d-i). Nonetheless, SWASH captured the variation and typical magnitude of the added mass and radiation damping for all degrees of freedom (i.e., the diagonal contributions of the radiation damping matrix). Discrepancies were largest for roll, for which the added mass $A_{\theta_x}$ was under predicted whereas the radiation damping $B_{\theta_x}$ was generally over predicted (especially for lower periods, Fig. \ref{fig:BEM_DifRad}g). Note that given the relatively small breadth of the vessel with respect to its length, roll hydrodynamic coefficients are at least an order of magnitude smaller compared to the other two rotational degrees of freedom (pitch $\theta_y$ and yaw $\theta_z$).

\begin{figure}[h!]
    \noindent
    \makebox[\textwidth]{\includegraphics{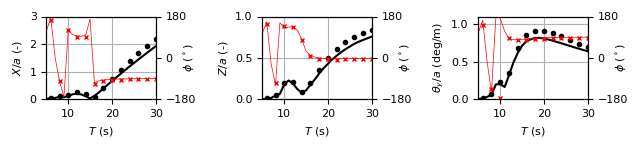}}
    \caption{Comparison between SWASH (markers) and BEM (lines) for the wave-induced response of a ship-shaped vessel moored in open water subject to monochromatic head waves. The black line and circles indicate the amplitude of the respective motion (left axis) and the red line and crosses indicate the phase difference with the incident wave signal (right axis).}
    \label{fig:BEM_RAO}
\end{figure}

Consistent with the diffraction and radiation problem, SWASH captured the vessel motion RAO amplitudes and phasing in surge, heave and pitch for the considered range of monochromatic head waves (Fig. \ref{fig:BEM_RAO}). This includes the relatively small response at low wave periods, and the increased response towards longer wave periods. The model also recovered the increased response of all three motions near $T=10$ s.

\subsubsection{Long-crested irregular waves}\label{sec:TC-MSOW-LCIW}

\begin{figure}[h]
    \noindent
    \makebox[\textwidth]{\includegraphics{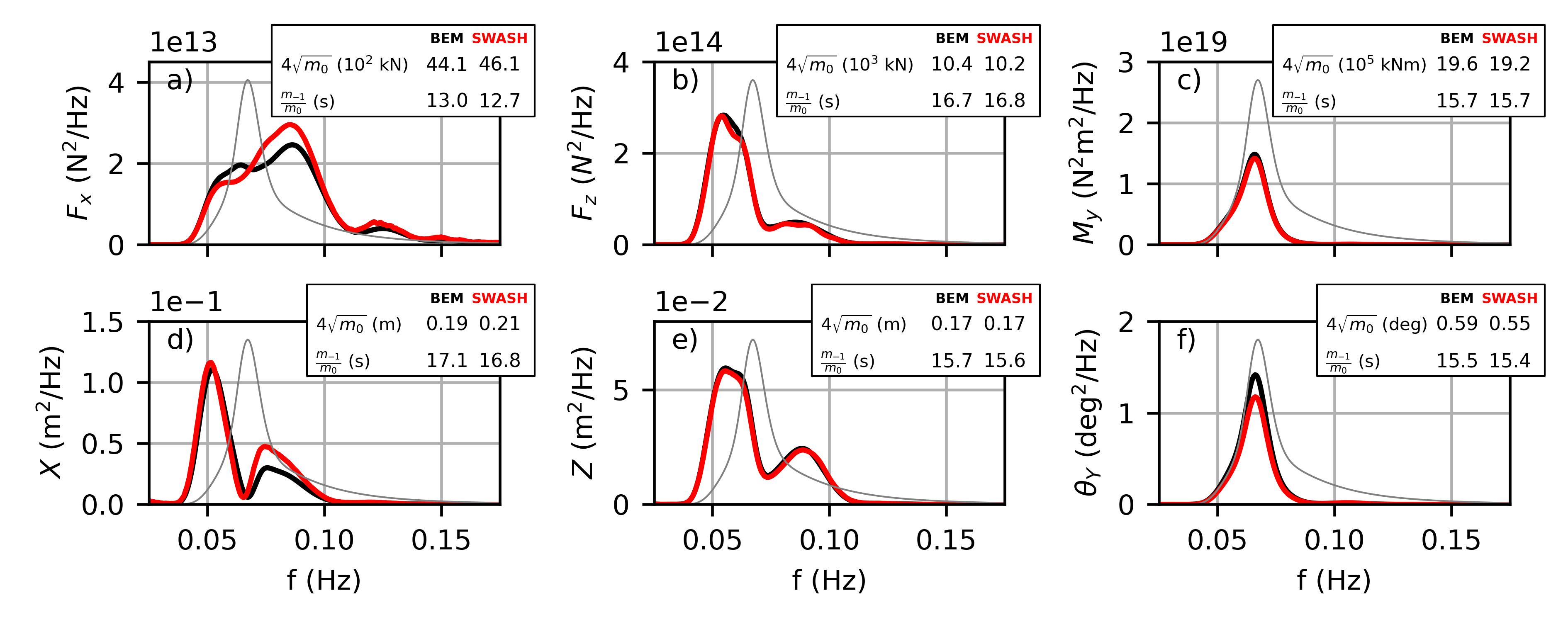}}
    \caption{Comparison between SWASH (red line) and BEM (black line) for the diffraction problem (top panels) and wave-induced response (bottom panels) of a ship-shaped vessel moored in open water subject to a long-crested sea-state (head waves). Bulk parameters based on spectral moments are included for each component in their respective panel. The thin gray line represents the incident wave spectrum.}
    \label{fig:BEM_DifRad_LC}
\end{figure}

Following the comparison for monochromatic waves, we consider a more realistic long-crested irregular wave field (JONSWAP spectral shape) with $H_s=1$ m and $T_p=15$ s. For the diffraction problem, a comparison between the power spectra (computed without smoothing) of the non-zero force and moment components shows that the SWASH predictions were in good agreement with the BEM solution (Fig. \ref{fig:BEM_DifRad_LC}a-c). The general agreement between SWASH and the BEM solution is confirmed by the bulk loads ($4\sqrt{\textrm{m}_0}$) and mean load periods ($\frac{\textrm{m}_{-1}}{\textrm{m}_0}$) measures (with maximum differences $<5\%$). The two models show a similar agreement for the wave-induced response in surge, heave and pitch (Fig. \ref{fig:BEM_DifRad_LC}d-e). SWASH captured the frequency variation and magnitude of all three motions, as confirmed by the bulk load parameters which differed by $<10\%$.

\subsubsection{Short-crested irregular waves}\label{sec:TC-MSOW-SCIW}
As a final comparison for the linear wave-induced response, we consider a short-crested sea state with a JONSWAP shape and a $\cos^m{\left(\theta\right)}$ directional distribution. Similar to the previous test cases, the sea state had a significant wave height of $H_s=1$ m and a peak period of $T_p=15$ s. The power of the directional distribution was $m=4$ which results in a one-sided directional spreading of $\sigma_\theta\approx25^\circ$. To allow for a direct comparison with SWASH, the BEM based solutions were computed based on the target wave component with which the numerical wavemaker in SWASH was forced. Power spectra for all load and motion components were computed using Welch's method with fifteen $50\%$ overlapping windows.

\begin{figure}[b!]
    \noindent
    \makebox[\textwidth]{\includegraphics{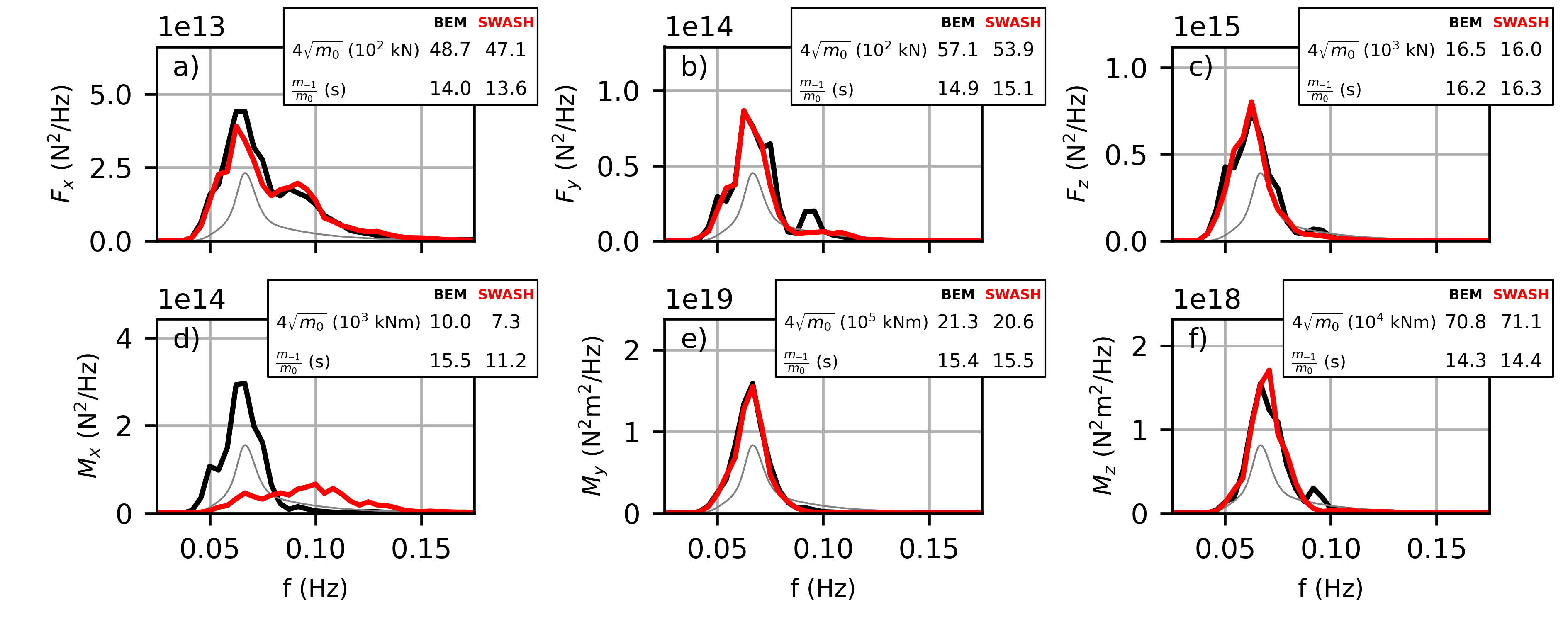}}
    \caption{Comparison between SWASH (red line) and BEM (black line) for the diffraction problem of a ship-shaped vessel moored in open water subject to a short-crested sea-state. Bulk parameters based on spectral moments are included for each component in their respective panel. The thin gray line represents the incident wave spectrum.}
    \label{fig:BEM_Dsc}
\end{figure}

\begin{figure}[h!]
    \noindent
    \makebox[\textwidth]{\includegraphics{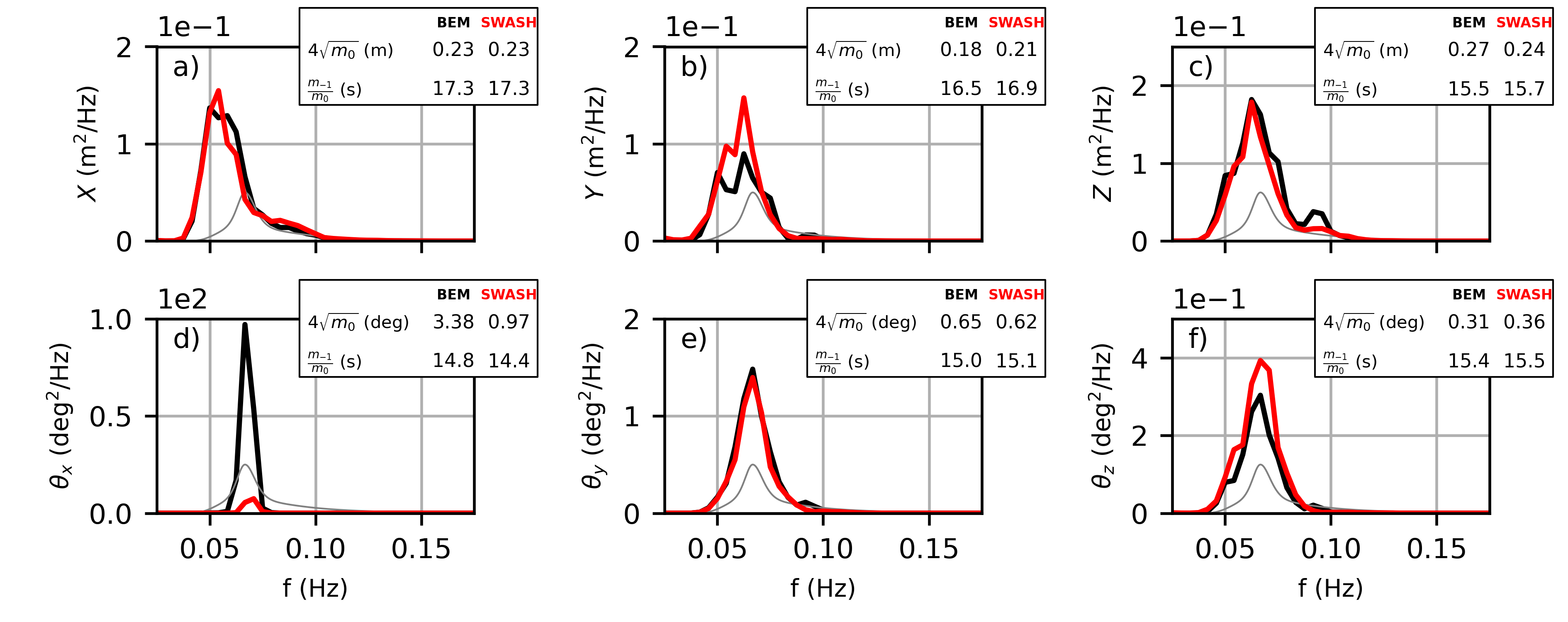}}
    \caption{Comparison between SWASH (red line) and BEM (black line) for the wave-induced response of a ship-shaped vessel moored in open water subject to a short-crested JONSWAP sea-state (head waves). Bulk parameters based on spectral moments are included for each component in their respective panel. The thin gray line represents the incident wave spectrum.}
    \label{fig:BEM_Rsc}
\end{figure}

SWASH captured the typical frequency dependence of the excitation forces and moments (Fig. \ref{fig:BEM_Dsc}), except for the roll excitation moment $M_x$ (Fig. \ref{fig:BEM_Dsc}d). Accordingly, both the bulk excitation force and period measures were in good agreement (discrepancies $<6\%$) for the three excitation forces (Fig. \ref{fig:BEM_Dsc}a-c) and the pitch and yaw moments (Fig. \ref{fig:BEM_Dsc}e-f). For the roll excitation moment, SWASH predicted a broader frequency distribution compared to the BEM solution, with smaller spectral levels near the peak frequency of the incident waves ($f_p$) and larger spectral levels at frequencies larger than $f_p$. The roll moment may be decomposed into contributions from the heave and sway exciting forces. For this hull and sea state, we found that the two individual roll contributions are of opposing sign with a magnitude that is several times larger than the resulting roll moment (i.e., $4\sqrt{m_0}\approx5\times10^7$ kNm for $M_{x}$ from ${F_z}$ and ${F_y}$ versus $4\sqrt{m_0}\approx 1\times10^7$ kNm for the total $M_x$). As a result, $M_x$ is sensitive to minor deviations in the underlying contributors. This sensitivity is confirmed by a grid sensitivity test (\ref{sec:A-GridSensitivity}), which shows that the SWASH predicted roll moment is sensitive to the horizontal grid resolution, with improving agreement for finer horizontal grid resolutions. In contrast, the roll excitation moment was not found to be sensitive to the vertical grid resolution.

Consistent with the excitation forces and moments, the wave-induced motions of the moored vessel were generally in good agreement (Fig. \ref{fig:BEM_Rsc}). Discrepancies were largest for the roll motion $\theta_x$, for which the BEM model predicted a peaked response at the resonance frequency ($f\approx0.07$ Hz). Although SWASH did capture a peaked response at the same frequency, the spectral levels of the roll motion were smaller. As a result, the SWASH predicted roll period was in good agreement with the BEM result, whereas the SWASH predicted bulk roll motion was a factor $3-4$ smaller compared to the BEM result (Fig. \ref{fig:BEM_Rsc}d). Similar to the roll excitation moment, the roll response was found to be sensitive to the horizontal grid resolution of SWASH (\ref{sec:A-GridSensitivity}). For the three translations (surge, sway and heave) and two remaining rotations (pitch and yaw), SWASH captured the frequency dependence and magnitude well, as indicated by relatively small differences ($<15\%$ between the bulk motion parameters (Fig. \ref{fig:BEM_Rsc})).

\subsection{Moored ship vessel in open water -- second order response}\label{sec:TC-MSOW-O2}
As a final test case, we consider the second-order response of a ship that is moored in open water. We consider the same set-up as in Section \ref{sec:TC-MSOW-LCIW}, with a ship moored at 28.6 m depth that is subject to a long-crested irregular wave field with $H_s= 1$m and $T_p=15$ s. Here, we will focus on the low-frequency second-order response of the moored ship that arises due to non-linearity of the incident wave field (i.e., the presence of low-frequency infragravity waves that are excited by difference interactions between the primary components), scattering of these waves and second-order contributions to the hydrodynamic forces due to (nonlinear) products of first-order quantities \citep[e.g.,][]{vanderMolen2008}. The nonlinear contributions are intrinsically accounted for by SWASH, but need to be explicitly accounted for in the BEM model.

\subsubsection{BEM model set-up}
To solve for the second-order difference-frequency forces and responses we employ the BEM code DIFFRACT, which has been extensively validated for second-order computations \citep[e.g.][]{Zang2006SecondWaves,Sun2013NonlinearColumn}. For the complete second-order computation, the mean free surface must be meshed out to a given radius -- here taken to be 500 m -- beyond which the forcing due to the nonlinear free surface boundary condition is represented by simplified forms in two annular regions, with the outer region being the far-field asymptotic region. In the second-order problem, first-order body motions contribute to second-order forces, so that the second-order hydrodynamic force on the fixed vessel is not the same as that on the floating vessel.  However, in the present case these differences are minor. DIFFRACT computes elements in the quadratic transfer function (QTF) matrix through bichromatic (pair-wise) computations.  The QTF matrix is sufficiently smooth to be interpolated, and the force/response spectra computed directly in the frequency domain. In this case an $8 \times 8$ QTF matrix was sufficient, though finer frequency spacing was employed as a check. The correction to the second-order heave force detailed by \cite{EatockTaylor1989IsLoads,Chen2006TheWaves} must be included to give good comparisons at low frequencies.

\subsubsection{SWASH methodology}
We used a similar SWASH model setup as in Section \ref{sec:TC-MSOW-SCIW} with an extended domain in $y-$direction (with a total length of 5016 m) to minimise the influence of side-wall reflections. Waves were generated by a weakly reflective wavemaker with a second-order correction for difference interactions to account for bound infragravity waves at the boundary \citep[e.g.,][]{Rijnsdorp2014}. With this wavemaker, the generation of spurious free waves is suppressed at sub-harmonic frequencies. Although spurious free waves were excited at super-harmonic frequencies \citep[e.g.,][]{Fiedler2019TheModeling}, they had a negligible effect on the wave-excitation loads and body motions.

To distinguish between the first- and second-order wave motion, excitation loads and body motions predicted by SWASH; we employ the phase separation method  \citep[e.g.,][]{Fitzgerald2014PhaseColumn}. This methodology assumes that the nonlinear wave-driven processes can be described by a Stokes-type perturbation expansion, and has been successfully applied to study both nonlinear wave processes \citep[e.g.,][]{Orszaghova2014ImportanceOvertopping,Whittaker2017OptimisationBeach,Zhao2017GapGroups} and nonlinear wave dynamics of fixed and moving structures \citep[e.g.,][]{Fitzgerald2014PhaseColumn,Chen2021ExtremeGroups,Orszaghova2021Wave-andTurbine}. To separate the primary and second-order contributions with this methodology, we ran a simulation with two different wavemaker signals that are out of phase (i.e., all wave frequencies of the second wavemaker signal are phase shifted by $180^\circ$ relative to the first wavemaker signal). Ignoring third and higher order components (which were negligibly small in the simulations of this work), this methodology allows estimations of the first-order (subscript L) and second-order (subscript NL) contributions to time-signals of the surface elevation, excitation loads (forces and moments) and body response as,

\begin{align*}
    X_\textrm{L} = \frac{1}{2} ( X_{0} - X_{180} ),\\
    X_\textrm{NL} = \frac{1}{2} ( X_{0} + X_{180} ),
\end{align*}
in which $X$ is the time-signal of interest, and the subscript indicates the phase shift (in degrees) of the wave components in the corresponding simulation.

\subsubsection{Results}

\begin{figure}[t]
    \noindent
    \makebox[\textwidth]{\includegraphics{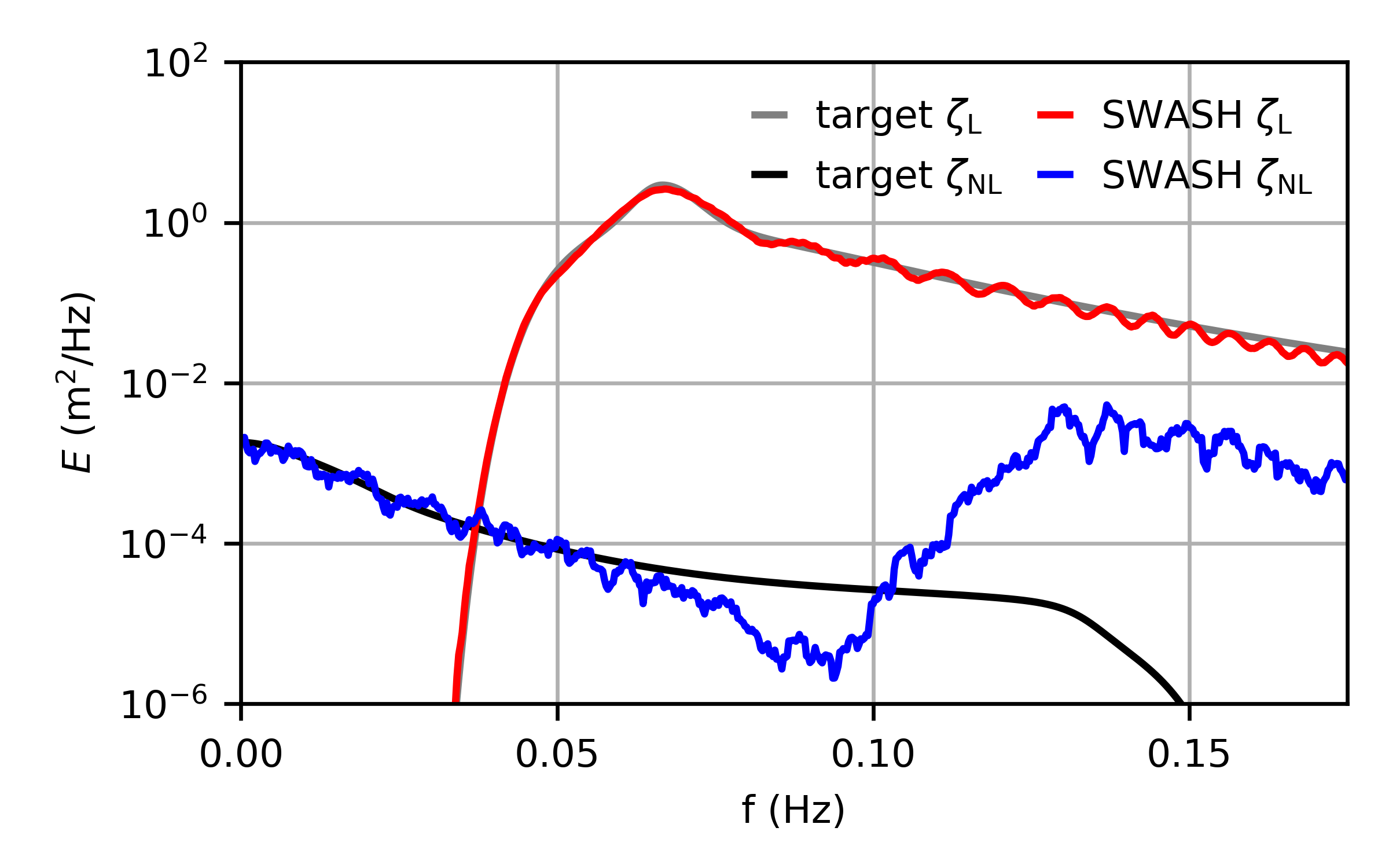}}
    \caption{Wave spectra of the linear and nonlinear surface elevation signals. The black and gray line indicate the target first-order (subscript L) and second-order (subscript NL) signal with which the wavemaker is forced, and the blue and red line indicate the first-order and second-order surface elevation signals inside the SWASH domain (obtained with the phase-separation methodology).}
    \label{fig:BEM_WSO2}
\end{figure}

The spectra of the linear and nonlinear surface elevation signals ($\zeta_\textrm{L}$ and $\zeta_\textrm{NL}$, respectively) inside the SWASH domain confirm that the generated wave field compares well with the target wave spectrum with which the wavemaker was forced (Fig. \ref{fig:BEM_WSO2}) at the sub-harmonic frequencies ($f<0.05$ Hz). The nonlinear energy spectrum starts to deviate from the target spectrum for $f>0.05$ Hz with larger spectral energy levels around $2 f_p$ associated with sum interactions between primary wave components. These second-order super-harmonic frequencies are not included in the wavemaker signal resulting in additional spurious free waves at $f>f_p$. However, the energy levels of second-order wave components are several orders of magnitude smaller than the energy levels of the primary (linear) wave components at the primary wave frequencies ($f>f_p/2$), indicating that the second-order super harmonic wave components did not significantly affect  the ship hydrodynamics.

\begin{figure}[h!]
    \noindent
    \makebox[\textwidth]{\includegraphics{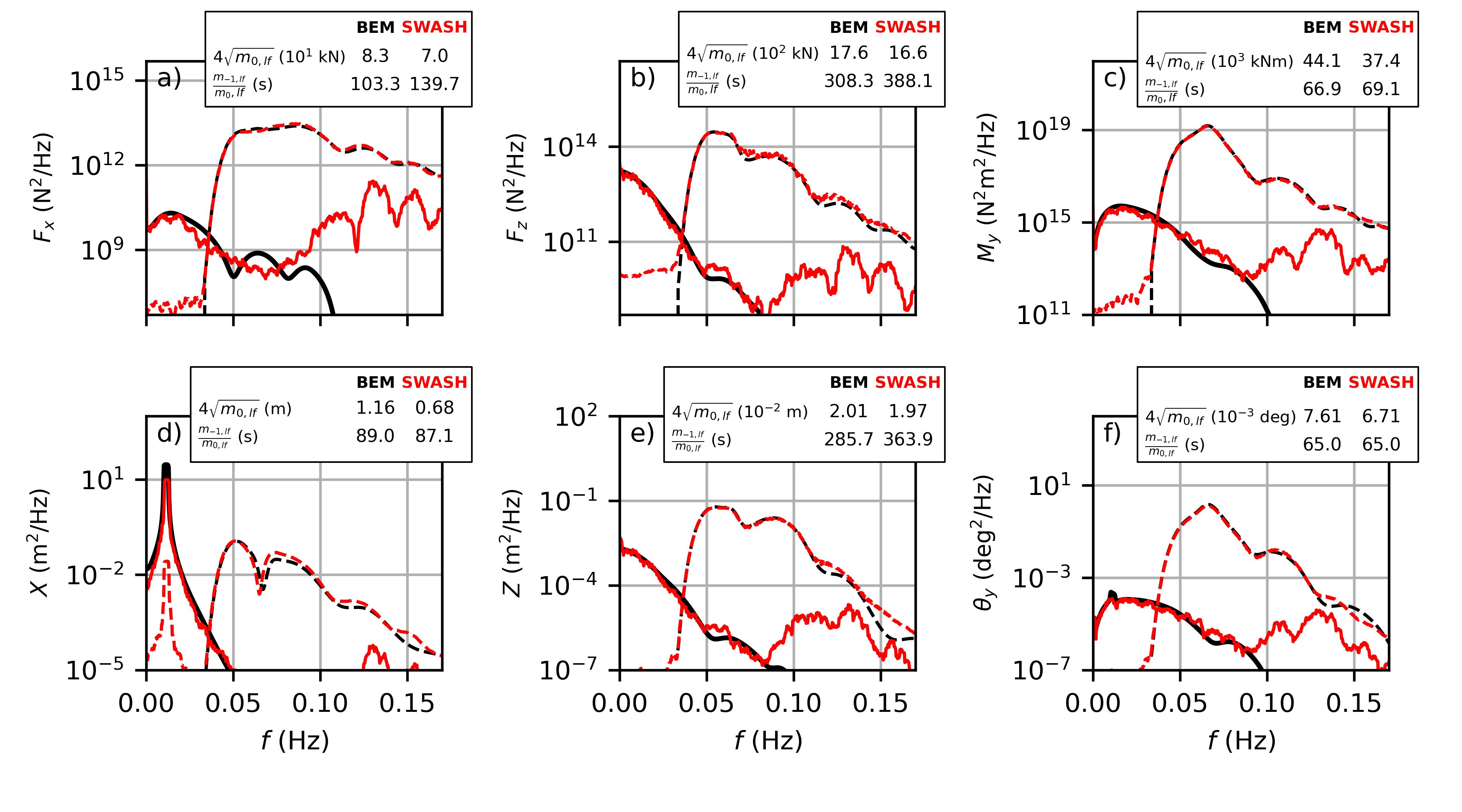}}
    \caption{Comparison between the first-order (dashed lines) and second-order (solid lines) results of SWASH (red lines) and BEM (black lines) for a ship-shaped vessel moored in open water that is subject to a long-crested JONSWAP sea-state (head waves). The top panels show the non-zero excitation forces (i.e., the diffraction problem) and the bottom panels show the non-zero body motions.  Bulk parameters based on low-frequency ($0<f<0.04$ Hz) spectral moments (indicated by subscript lf) of the second-order signals are included for each component in their respective panel.}
    \label{fig:BEM_DRO2}
\end{figure}

The second-order excitation loads were (at least) an order of magnitude smaller than the first-order loads for all three non-zero components (Fig. \ref{fig:BEM_DRO2}a-c). The nonlinear excitation load from SWASH agreed well with the second order BEM solution at the subharmonic frequencies ($f<0.05$ Hz) in terms of the spectral shape and spectral levels. The second-order contributions in both models exceeded the linear response for $f<0.04$ Hz. To quantify the agreement between the two models, we computed the bulk parameters  by integrating the second-order spectra over a low-frequency (lf) band where the second-order response exceeded the linear response ($f<0.04$ Hz). The bulk parameters ($4\sqrt{m_{0,\textrm{lf}}}$ and $\frac{m_{{-1,\textrm{lf}}}}{m_{0,\textrm{lf}}}$) of the two  models differed $<16\%$, confirming the good agreement between the SWASH and BEM model.

For the BEM results it is straightforward to separate the second-order Froude-Krylov force due to the incident bound waves, the additional force due to scattering of these waves (with an inhomogeneous free surface boundary condition), and quadratic forces due to products of first order terms.  As might be expected, BEM results confirmed that the Froude-Krylov terms dominated the second-order forces in heave and pitch, however, in surge the quadratic and Froude-Krylov terms were of similar magnitude.  The second-order forces due to scattering of the second order incident potential are small in the range $f<0.04$ Hz.

Both models predict a second-order vessel response in heave and pitch that is several orders of magnitude smaller compared to the linear response (Fig. \ref{fig:BEM_DRO2}e-f). The spectral shape and spectral levels, as well as the bulk parameters indicate that both models predict a similar second-order heave and pitch response (with differences in bulk parameters $<10\%$). In contrast with heave and pitch, both models predict a second-order surge response that is an order of magnitude larger compared to the linear response with a distinctive peak at a resonant frequency of $f\approx0.01$ Hz (Fig. \ref{fig:BEM_DRO2}a). SWASH predicts a similar second-order surge response compared to the second-order BEM solution with a resonant response at $f\approx=0.01$ Hz. SWASH however underpredicts the BEM surge magnitude by about $40\%$. We further note that the model captured particular features of the motion spectra, such as the bump in the pitch spectrum at the surge natural frequency that is associated with pitch-surge coupling (Fig. \ref{fig:BEM_DRO2}f).


\section{Discussion and conclusions}\label{sec:Conclusions}
In this paper we propose a new numerical model based on the non-hydrostatic framework to predict the wave-induced response of a floating structure that is moored in a coastal or harbour region. The methodology is implemented in the open-source non-hydrostatic model SWASH, an efficient tool to simulate the nonlinear evolution of waves over variable bottom topography. In this work we have extended SWASH with a solution to the rigid body equations (governing the motions of the floating structure) that is tightly coupled to the hydrodynamic equations (governing the water motion) to simulate the wave-induced response of a moored floating structure.

The model was validated for two test cases, 1) a moored floating cylinder that is representative of a wave-energy-converter (with dimensions that are small relative to the wave length), and 2) a moored ship-shaped vessel (with dimensions approximately equal to the wave length). We tested the model for the diffraction problem, radiation problem, and the wave-induced response of the two floating structures. We considered a range of sea-states, varying from monochromatic waves up to more realistic short-crested sea states. First, we compared model predictions with a benchmark potential flow solution for the (first-order) primary wave-components (sea and swell waves). Predictions of the excitation forces (diffraction problem) and hydrodynamic coefficients (radiation problem) were compared to a semi-analytical and a BEM solution to the linear (first-order) potential flow problem. The predicted wave-induced response was similar compared to a frequency domain solution using linear potential flow theory. Model predictions were in good agreement with the linear reference solutions for both floating structures for the variety of wave conditions that were considered.

The model accounts for the fully nonlinear kinematic boundary condition at the free surface, combined with linearized kinematic boundary conditions at the hull of the structure (as the flow grid is not adapted for the motions of the floating structure). The latter formally restricts the method to small motions of the structure with respect to the flow grid. When this assumption is satisfied, the model intrinsically accounts for various non-linear effects that may contribute to the wave-induced response of a moored floating structure. This includes non-linear Froude-Krylov contributions from higher-order wave components (e.g., due to bound infragravity waves), quadratic terms (e.g., due to the wetting of the hull), and the scattering of higher-order wave components. We confirmed this by comparing the model for the second-order (difference frequency) hydrodynamic loads and motions of a ship-shaped vessel moored in open water. SWASH predictions were in satisfactory agreement with a second-order BEM solution for both the low-frequency excitation loads and response of the moored structure. In particular, the SWASH model was able to capture the resonant low-frequency response in surge.

The inherent inclusion of various nonlinear effects sets the SWASH model apart from BEM models based on a perturbation expansion approach, which require the application of higher-order formulations \citep[e.g.,][]{You2015} or coupling to nonlinear wave propagation models such as Boussinesq-type models \citep[e.g.,][]{Bingham2000,vanderMolen2008} to account for nonlinear effects. Perturbation-type BEM models are inherently limited to weakly nonlinear sea-states as the assumptions of the perturbation approach may be violated in extreme sea states. Fully nonlinear potential flow models \citep[e.g.,][]{Ma2009,Yan2019ComparativeStructure,Hu2020InvestigationSolvers}, e.g. using the Finite Element Method (FEM),  can also intrinsically account for nonlinear wave-structure interactions but are typically not applied in extensive coastal regions. By solving the fully nonlinear kinematic boundary condition at the water surface (similar to FEM), the proposed model is able to resolve fully nonlinear waves and we expect that it provides enhanced predictions over BEM based models for more extreme sea-states.  In addition, as a wave-flow model, SWASH has the ability to simulate currents, either large scale or wave-driven, with additional effects on the wave field and vessel.

Mooring lines introduce an additional source of nonlinearity, even when motions of a moored structure are small. In this work, we modelled mooring lines as linear spring-dampers. Such a simple representation of a mooring line is not necessarily sufficiently accurate to capture the wave-induced response of real moored structures \citep[e.g.,][]{Bingham2000}. Although not included in this work, the proposed model could be extended with more accurate representations of mooring lines. For example, with future work the model can be coupled to dynamic mooring line models \citep[e.g.,][]{Palm2016CoupledValidation,Hall2015ValidationData} to provide a more sophisticated representation of the mooring line dynamics.

The major limitation of the proposed model is the relatively coarse schematisation of the hull of the floating structure due to the use of a single valued function for the vertical fluid boundaries (i.e., the free-surface and the hull, see Section \ref{sec:NumericalMethodology}). The model cannot represent complex hull features such as a bulbous bow, in contrast with existing methodologies that provide a more detailed schematisation of the hull (e.g., BEM models). The agreement between SWASH and reference solutions for two relatively simple hulls (Section \ref{sec:TC-FCOW}-\ref{sec:TC-MSOW}) indicates that the model will be able to capture the net hydrodynamic loads and body motions of a structure when the influence of complex hull features on the hydrodynamics is relatively small.

A further limitation is that the model is not able to capture details of flow separation from body surfaces or turbulent flow fields due to a relatively coarse spatial resolution. Similar to potential flow models, the influence of viscous damping on the body motions will need to be explicitly included through appropriate parametrisations \citep[e.g.,][]{Bingham2000}. As is well known, such viscous effects will be significant for low-frequency surge response and roll motion, for example.  The low-frequency horizontal motions may also require additional damping to represent wave-drift damping. If representing viscous effects more directly is desired, Computational Fluid Dynamics (CFD) models including appropriate turbulence models are likely better suited to simulate the wave-structure interactions \citep[e.g.,][]{Hadzic2005,Wilson2006,Chen2021ExtremeGroups}, though at far greater computational cost.

With the aforementioned modelling assumptions, we believe that the model is suited to simulate the response of a range of floating structures that are moored in both sheltered regions and in open water of restricted depth. As long as the motions of the structure are relatively small (which is generally a reasonable assumption for moored structures), the model does not impose any restrictions on the wave nonlinearity and is thus in principle able to deal with energetic sea states as long as viscous effects can be parametrized appropriately. For all considered test cases in this work, the model captured the wave-structure interactions and wave-induced response with a coarse vertical resolution. Such coarse resolutions allow for model applications at the scale of a realistic harbour or coastal region ($\approx \mathcal{O}\left(1-10\right)$ km$^2$). The computational requirements of the proposed methodology is in fact comparable to wave-propagation models based on Boussinesq-type equations and the non-hydrostatic approach. For example, the simulations considered in this work took about twice as long as conventional SWASH simulations excluding the floating structure. This work thereby presents a new modelling tool to seamlessly simulate the nonlinear evolution of waves from deep to shallow water over complex bottom topography and the wave-induced response of a floating structure that is moored in coastal waters.

\section*{Acknowledgements}
The computational part of this research was supported by resources provided by the Pawsey Supercomputing Centre with funding from the Australian Government and the Government of Western Australia. The authors thank Nataliia Sergiienko for making a Matlab code with a semi-analytical solution to the linearized potential flow equations for emerged and submerged cylinders publicly available via ResearchGate. The extension of the SWASH model will be included in an upcoming release of the model, and will be available from \url{https://swash.sourceforge.io} under the GNU-GPL user license.

\appendix
\setcounter{figure}{0}

\section{Grid sensitivity}\label{sec:A-GridSensitivity}

We conducted a grid sensitivity test to understand the influence of the grid resolution on the SWASH predictions using the test case that considers a ship moored in open water that is subject to a short-crested irregular sea-state (Section \ref{sec:TC-MSOW-SCIW}). In this test case, significant discrepancies were observed between the SWASH and BEM predicted roll exciting moment and roll response (Fig. \ref{fig:BEM_Dsc}d-\ref{fig:BEM_Rsc}d).
We repeated the same simulation but with a finer horizontal and vertical grid resolution. We considered a horizontal grid resolution of $\Delta x=4$ m (the original resolution), $\Delta x=2$ m  and $\Delta x=1$ m, with a constant vertical resolution of two layers. To test the sensitivity to the vertical resolution, 2 layers (the original resolution), 3 layers and 5 layers were considered, with a constant horizontal grid resolution of $\Delta x=2$ m.

\begin{figure}[h]
    \noindent
    \makebox[\textwidth]{\includegraphics{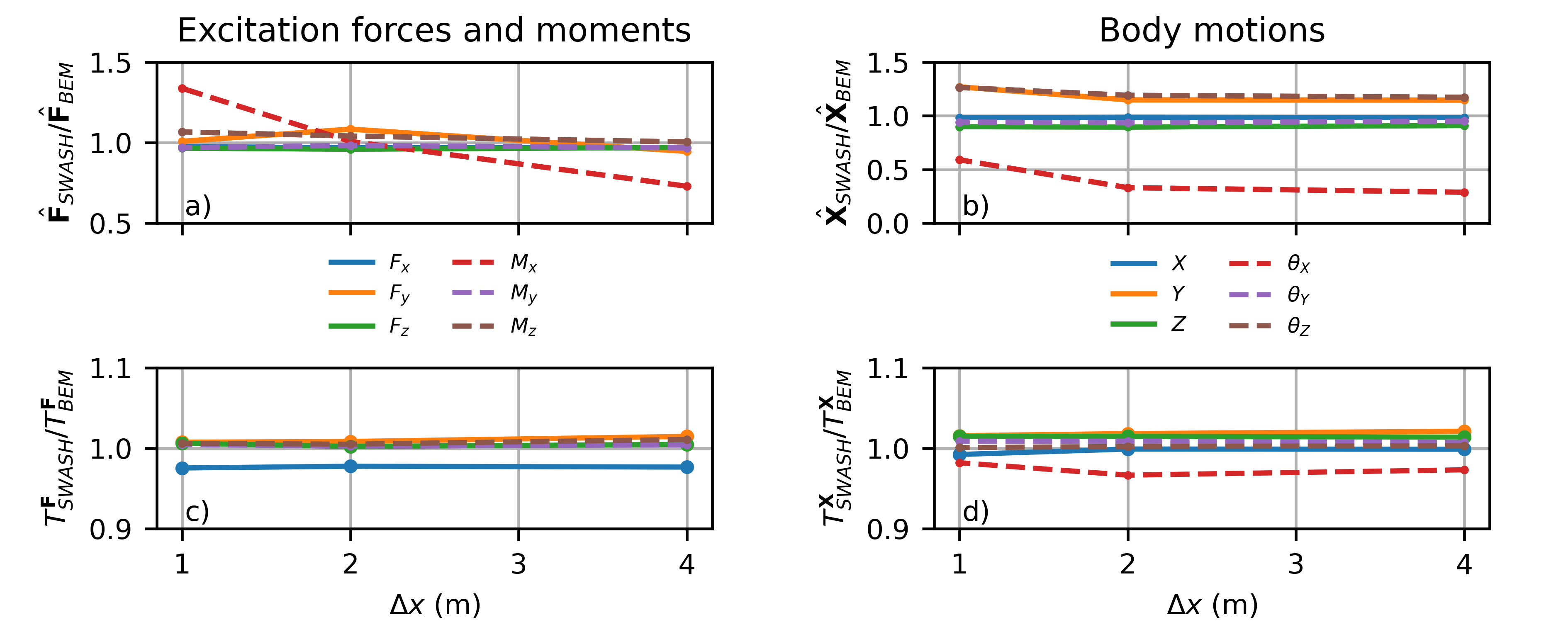}}
    \caption{Bulk parameters of the excitation loads ($\hat{\mathbf{F}}=4\sqrt{m_0}$ and $T^{\mathbf{F}}=m_{-1}/m_0$), panel a-b) and of the body motions ($\hat{\mathbf{X}}=4\sqrt{m_0}$ and $T^{\mathbf{X}}=\frac{m_{-1}}{m_0}$, panel c-d) as a function of the horizontal grid resolution $\Delta x$. The bulk parameters from SWASH (subscript SWASH) are normalized by the results from the BEM simulations (subscript BEM). Colours indicate the different load and motion component (as indicated by the legend between the panels).}
    \label{fig:BEM-Sens-dx}
\end{figure}

\begin{figure}[h]
    \noindent
    \makebox[\textwidth]{\includegraphics{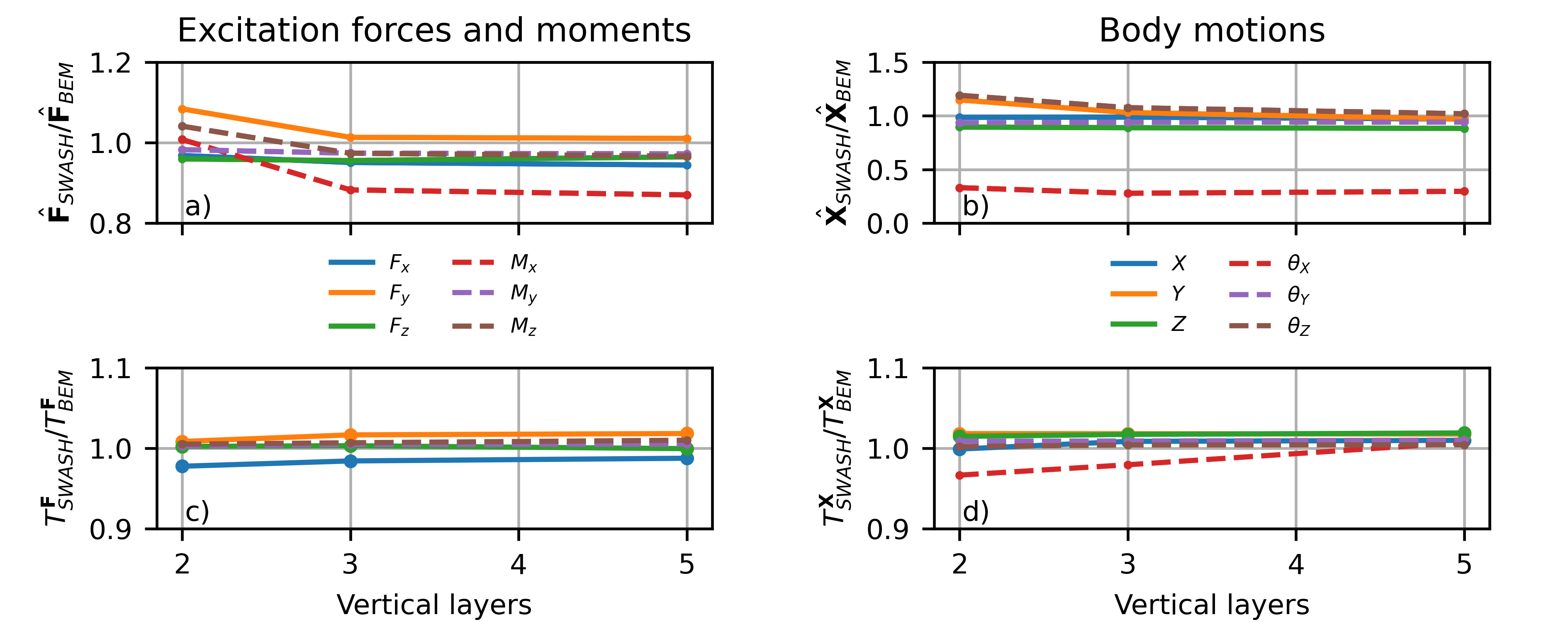}}
    \caption{Bulk parameters of the excitation loads ($\hat{\mathbf{F}}=4\sqrt{m_0}$ and $T^{\mathbf{F}}=m_{-1}/m_0$), panel a-b) and of the body motions ($\hat{\mathbf{X}}=4\sqrt{m_0}$ and $T^{\mathbf{X}}=\frac{m_{-1}}{m_0}$, panel c-d) as a function of the vertical grid resolution (number of layers). The bulk parameters from SWASH (subscript SWASH) are normalized by the results from the BEM simulations (subscript BEM). Colours indicate the different load and motion component (as indicated by the legend between the panels).}
    \label{fig:BEM-Sens-Nv}
\end{figure}

To quantify the sensitivity of the results to the grid resolution, we compare the bulk parameters ($4\sqrt{m_0}$ and $\frac{m_{-1}}{m_{0}}$) for all 6 excitation loads (forces and moments) and all 6 body motions as a function of the horizontal grid resolution (Fig. \ref{fig:BEM-Sens-dx}) and the vertical grid resolution (Fig. \ref{fig:BEM-Sens-Nv}). The excitation loads and body motions are relatively insensitive to the vertical grid resolution, with relatively small changes in both bulk parameters (less than $\approx15\%$ relative to the BEM solution) when increasing the number of vertical layers (Fig. \ref{fig:BEM-Sens-Nv}). Changes in the period measure ($\frac{m_{-1}}{m_{0}}$) were especially small, indicating that the vertical resolution did not significantly affect the spectral shapes. Most of the bulk parameters were also relatively insensitive to the horizontal grid resolution, except for the roll moment and roll motion (Fig. \ref{fig:BEM-Sens-dx}). In particular, the magnitude of the roll moment and roll motion increases for a finer grid resolution (Fig. \ref{fig:BEM-Sens-dx}a-b), although the period measure of $M_x$ and $\theta_x$ is less affected (Fig. \ref{fig:BEM-Sens-dx}c-d). This suggest that the spectral levels of $M_x$ and $\theta_x$ increase for finer horizontal grid resolutions, but that the spectral shape remains roughly intact.

\begin{figure}[h]
    \noindent
    \makebox[\textwidth]{\includegraphics{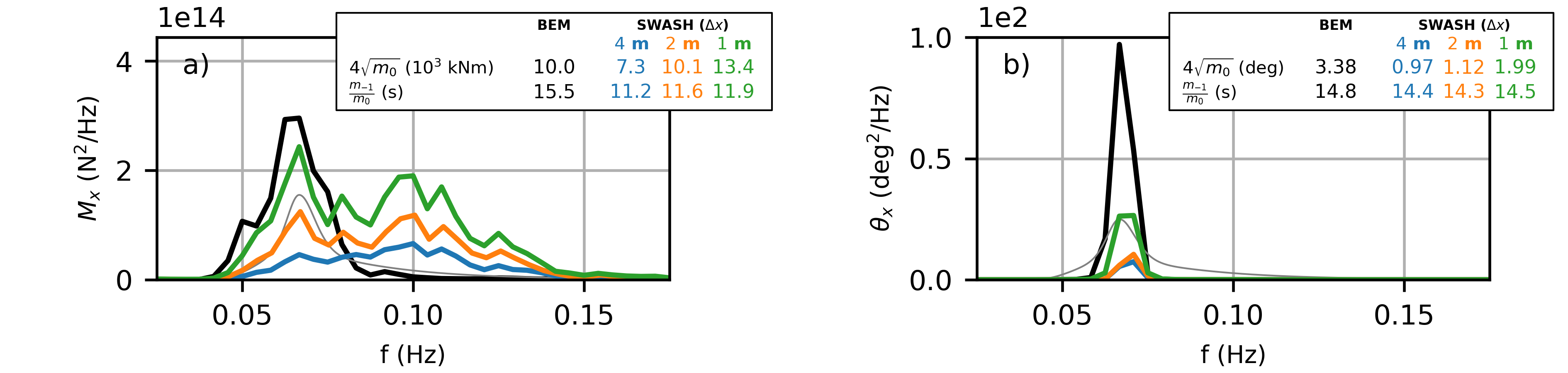}}
    \caption{Sensitivity of the SWASH predictions (coloured lines) to the horizontal grid resolution ($\Delta x=1-4$ m, with a constant vertical resolution of 2 layers) for the roll excitation moment (Diffraction problem) and the roll motion (wave-induced response) of a ship-shaped vessel moored in open water subject to a short-crested JONSWAP sea-state (head waves). The black line indicates the reference BEM solution, and the thin gray line indicates the incident wave spectrum. Bulk parameters based on spectral moments are included in both panels. The color of the SWASH results reflects the horizontal grid resolution, as indicated by the panel with bulk parameters.}
    \label{fig:BEM_Mx-Sens-dx}
\end{figure}

To gain further insight in the effect of the horizontal grid resolution on the spectral shape of the roll moment and response, Fig. \ref{fig:BEM_Mx-Sens-dx} shows the spectra of $M_x$ and $\theta_x$ for the considered $\Delta x$. For increasingly fine grid resolutions, $M_x$ indeed increased over a wide spectral range and better matched the BEM solution near the peak incident wave frequency $f_p$ (Fig. \ref{fig:BEM_Mx-Sens-dx}a). However, $M_x$ was increasingly over predicted for $f>f_p$, resulting in an over prediction of the bulk roll moment for the finest grid resolution whereas the period measure changed marginally (Fig. \ref{fig:BEM-Sens-dx}c and \ref{fig:BEM_Mx-Sens-dx}a). The SWASH predicted roll response showed a peaked response at $f\approx0.07$ Hz, in accordance with the BEM solution (Fig. \ref{fig:BEM_Mx-Sens-dx}b). For increasingly fine horizontal grid resolutions, SWASH better captured the magnitude of the roll motion, although the bulk roll motion still under predicted the BEM bulk roll motion by about $40\%$ (Fig. \ref{fig:BEM_Mx-Sens-dx}b).

\clearpage
\bibliographystyle{elsarticle-harv}
\bibliography{manuscript.bbl}

\end{document}